\documentclass[
  reprint,
  amsmath,
  amssymb,
  aps,
  pre,
  floatfix
]{revtex4-2}
\usepackage{graphicx,amsthm}
\usepackage{dcolumn}
\usepackage{bm}
\usepackage{xcolor}
\usepackage{comment}

\newtheorem{theorem}{Theorem}[section]
\newtheorem{lemma}[theorem]{Lemma}

\newtheorem{result}{Result}

\theoremstyle{definition}

\newtheorem{remark}[theorem]{Remark}

\DeclareMathOperator{\BR}{BR}

\newcommand{\E}{\mathbb{E}}

\newcommand{\1}{\mathbf{1}}

\newcommand{\T}{\mathsf{T}}

\begin{document}

\preprint{APS/123-QED}

\title{Stochastic Evolutionary Stability under Random Payoffs: Beyond First-Order Invasion Tests
}

\author{Dhaker Kroumi}
\email{dhaker.kroumi@kfupm.edu.sa}
 \affiliation{Department of Mathematics, King Fahd University of Petroleum and Minerals, Dhahran, S.A.}



\begin{abstract}
Environmental fluctuations can change invasion resistance beyond predictions based on mean fitness. In this paper, we consider competition between a resident and a rare mutant in an infinite, well-mixed population with discrete generations. Each phenotype uses a fixed pure or mixed strategy in a symmetric matrix game, and the fitness matrices are independent and identically distributed across generations. We complete the classification of stochastic Nash equilibria in terms of their best replies and determine which equilibria are stochastically evolutionarily stable. We show that an equilibrium always exists and that its best replies are exactly those strategies whose realized fitness against the resident equals the resident's own almost surely. To characterize stability, we resolve the case in which both the invasion exponent and the first frequency-dependent coefficient vanish. In this case, the second-order fitness coefficient and environmental fluctuations distinguish local attraction, almost-sure local escape, and exact neutrality. For each attracting regime, we determine the rate at which mutant frequency tends to zero on trajectories that remain sufficiently close to resident fixation: exponential decay, inverse-time decay, or slower. For tied mutants that escape when the first frequency-dependent coefficient vanishes, we identify the threshold separating finite from infinite mean exit times.
\end{abstract}

\maketitle
\section{Introduction}\label{sec:intro}
Changes in the environment affect how many offspring individuals produce, and so they can decide whether a rare mutant invades a resident population \cite{LandeEngenSaether2003}. Since growth multiplies from one generation to the next, what counts in the long run is the average of the logarithm of fitness, not the average fitness itself \cite{LewontinCohen1969}. In evolutionary games, fitness also depends on how common each strategy is \cite{HofbauerSigmund1998}. Whether a mutant can invade depends on both effects together. Broom \cite{Broom2005}, for example, studied how evolutionary stability changes when the payoff depends on when it takes place. Studying these two effects together extends the classical theory of evolutionary stability \cite{MaynardSmithPrice1973,MaynardSmith1982}. The hardest case is when the usual measures of mutant growth are zero, because they cannot tell us whether the mutant invades.

Stochastic local stability under fluctuating selection was first studied in population genetics \cite{KarlinLiberman1974,KarlinLiberman1975}. Zheng \textit{et al.}\ \cite{ZhengLiLessardTao2017} later brought this idea to games with random payoffs and defined stochastic evolutionary stability (SES) for two competing phenotypes. Their test looks first at the expected logarithmic relative fitness of a very rare mutant. When the two phenotypes earn the same realized fitness against the resident, this test is not enough, so they also look at the first change in expected logarithmic relative fitness as the mutant becomes more common. Feng \textit{et al.}\ \cite{FengMatrix2022} extended this approach to games with several phenotypes, and treated competition between two fixed mixed strategies through their four pairwise fitnesses. Building on this work, Li \textit{et al.}\ \cite{LiNoiseSNE2025} defined a stochastic Nash equilibrium (SNE) using geometric mean fitness. For games with two pure strategies, they proved that an SNE exists, that every SES is an SNE, and that a strong SNE is always SES, while a weak SNE may not be. Zheng \textit{et al.} \cite{ZhengTaoLi2025} have studied games with three pure strategies, including a cyclic game.

The works above study one particular dynamics, the exact recursion for an infinite population with discrete generations. The effect of environmental variation, however, depends on the population dynamics. For stochastic replicator equations, earlier studies examined the stability of equilibria and whether the dynamics are recurrent or transient \cite{FosterYoung1990,FudenbergHarris1992,Imhof2005,HofbauerImhof2009}, including when payoff fluctuations are correlated \cite{FengReplicator2022}. In these models, environmental noise can either stabilize an interior equilibrium or destabilize it \cite{ZhengLiLessardTao2018,FengRPS2022}, depending on the game. Fluctuating payoffs can also allow coexistence even when the arithmetic mean payoffs are unequal \cite{StollmeierNagler2018}. Under weak selection, by contrast, the leading stability conditions may depend mainly on the mean payoff matrix \cite{LiWeakSelection2019,LiJTB2023}, and similar mean payoff criteria appear under stochastic imitation \cite{WangImitation2024,WangTFT2026}. Finite populations add another layer, since environmental variation interacts with demographic sampling. This interaction affects both fixation probabilities \cite{AssafMobiliaRoberts2013,LiLessard2020,KroumiEtAl2021,KroumiMartinLessard2022} and the average abundance of strategies under recurrent mutation \cite{KroumiLessard2021,KroumiLessard2022}.

Unlike these approximations and finite-population models, we study the exact competition dynamics of two fixed, heritable phenotypes in an infinite population. Each phenotype may use a pure or mixed strategy. The fitness matrices are independent and identically distributed across generations, and their entries are uniformly positive and bounded. For any finite number of pure strategies, we first characterize the best replies to an SNE: a mutant ties with the resident if and only if both earn the same realized fitness against the resident, almost surely. For such a tied mutant, the invasion exponent is zero, so the question becomes whether the resident still resists it when the first frequency-dependent coefficient also vanishes. We show that three outcomes are possible: local asymptotic stability in probability, almost-sure local escape, and exact neutrality. Combined with the strict-sign criteria of Zheng \textit{et al.}\ and Feng \textit{et al.} \cite{ZhengLiLessardTao2017,FengMatrix2022}, this result completes the characterization of which SNE are SES, including residents on the boundary of the strategy simplex.

To show that these outcomes occur, we use games from the earlier literature. In a cyclic game of the form studied by Zheng \textit{et al.} \cite{ZhengTaoLi2025}, we find the fluctuation threshold at which the first frequency-dependent coefficient vanishes, and we show that the resident is still stable at this threshold. The matrix family of Li \textit{et al.}\ \cite[Example~A.2]{LiNoiseSNE2025} gives two further cases. In the first, which we extend to any finite number of pure strategies, the mean game is neutral, and yet every fixed distinct mutant goes extinct almost surely from every interior initial frequency. In the second, the resident resists both pure mutants, but mixed mutants arbitrarily close to it escape locally almost surely.

However, knowing that the resident is stable or unstable does not tell us how fast a mutant disappears or how long the population stays near an unstable fixation state. We  also study rates. As long as a trajectory stays in a small enough neighborhood of resident fixation, the mutant goes extinct in one of three ways: exponentially fast, like the inverse of time, or more slowly. For tied mutants that escape when the first frequency-dependent coefficient vanishes, we find when the expected exit time is finite or infinite. Our proofs use Lyapunov functions and stopping arguments \cite{BenaimSchreiber2019}, together with Lamperti's drift criteria on processes whose drift tends to zero \cite{Lamperti1960,DenisovKorshunovWachtel2025}. These distinctions matter when reading finite observations: a long stay near fixation does not prove stability, and a slow decline of the mutant does not prove neutrality.
\section{Model and invasion criteria}\label{sec:prelim}
Consider an infinite, well-mixed population with pairwise interactions and discrete, nonoverlapping generations. Individuals carry one of two distinct, heritable phenotypes, represented by the resident strategy \(x\) or the mutant strategy \(y\). Each strategy specifies an individual's probabilities of choosing among \(n\ge2\) pure actions and belongs to the simplex
\begin{equation}\label{eq:simplex_def}
\Delta_{n-1}:=\{x\in\mathbb R^n:x_i\ge0,\ \1^{\T}x=1\},
\end{equation}
where \(\1=(1,\ldots,1)^{\T}\), and all vectors are columns. During competition, these action probabilities remain fixed. Only the population frequencies of the two phenotypes can change.

The environment in generation \(t\) is represented by a random fitness matrix \(A(t)=(a_{ij}(t))_{i,j=1}^{n}\), where \(a_{ij}(t)\) denotes the reproductive fitness of an individual using pure action \(i\) against an opponent using pure action \(j\). Conditional on this matrix, \(y^{\T}A(t)x\) is the reproductive fitness of phenotype \(y\) against phenotype \(x\), averaged over their action choices. The game is symmetric in the sense that both competitors follow the same interaction rules, although \(A(t)\) need not be a symmetric matrix.

We impose three standing assumptions. \textbf{($\mathcal A1$)} The matrices $(A(t))_{t\ge0}$ are independent and identically distributed; entries within the same matrix may be dependent. \textbf{($\mathcal A2$)} There are two constants \(0<m<M<\infty\) such that
\begin{equation}\label{eq:uniform_bounds}
m\le a_{ij}(t)\le M
\end{equation}
a.s. for every \(i,j\in\{1,\ldots,n\}\) and \(t\ge0\). \textbf{($\mathcal A3$)} Residents and mutants experience the same realization of \(A(t)\) in each generation. Environmental variation is the only source of randomness in the population recursion.

A consequence of Assumption \textbf{($\mathcal A2$)} is that
\begin{equation}\label{eq:fitness_bounds}
m\le y^{\T}A(t)x\le M
\end{equation}
for all \(x,y\in\Delta_{n-1}\), a.s. These bounds ensure that logarithmic fitnesses are bounded and integrable. 

Now, let \(q_t\) be the frequency of the mutant phenotype at time \(t\geq0\). Then, the mutant and resident fitnesses are
\begin{subequations}\label{eq:realized_fitness}
\begin{align}
\pi_y(q_t,A(t))&=(1-q_t)y^{\T}A(t)x+q_ty^{\T}A(t)y,\label{eq:mutant_realized_fitness}\\
\pi_x(q_t,A(t))&=(1-q_t)x^{\T}A(t)x+q_tx^{\T}A(t)y.\label{eq:resident_realized_fitness}
\end{align}
\end{subequations}
We consider a reproduction model in which each individual contributes to the next generation in proportion to its reproductive fitness. More precisely, the mutant frequency evolves according to
\begin{equation}\label{eq:qrec}
q_{t+1}=\frac{q_t\pi_y(q_t,A(t))}{q_t\pi_y(q_t,A(t))+(1-q_t)\pi_x(q_t,A(t))}.
\end{equation}
This is the random fitness recursion studied in \cite{ZhengLiLessardTao2017,FengMatrix2022}.

Suppose that the initial frequency \(q_0\in(0,1)\) is deterministic. Positive fitnesses ensure that \(q_t\in(0,1)\) at every finite time $t\geq0$, so mutant extinction or fixation can occur only in the limit \(t\to\infty\).

To remove the normalization by mean population fitness in \eqref{eq:qrec}, we introduce the mutant odds
\begin{equation}\label{eq:mutant_odds}
z_t:=\frac{q_t}{1-q_t}.
\end{equation}
Define the logarithmic relative fitness by
\[
\ell(q,A):=\log\frac{\pi_y(q,A)}{\pi_x(q,A)}.
\]
Equation~\eqref{eq:qrec} then gives
\begin{subequations}
\begin{align}
z_{t+1}&=z_t\frac{\pi_y(q_t,A(t))}{\pi_x(q_t,A(t))},\label{eq:odds_recursion}\\
\log z_{t+1}-\log z_t&=\ell(q_t,A(t)).\label{eq:log_odds_recursion}
\end{align}
\end{subequations}
The logarithmic increments are therefore given by the logarithmic relative fitness, making cumulative growth a sum of these increments. Since \(z_t\to0\) if and only if \(q_t\to0\), extinction can be studied equivalently through the odds.

Let \(A\) be a random matrix with the common distribution of the environmental matrices \(A(t)\). For a fixed resident \(x\), define the expected logarithmic fitness of \(y\) against \(x\) by
\begin{equation}\label{eq:expected_log_fitness}
F(y,x):=\E\!\left[\log\bigl(y^{\T}Ax\bigr)\right].
\end{equation}
The invasion exponent is
\begin{equation}\label{eq:invasion_function}
Q(y,x):=F(y,x)-F(x,x)=\E[\ell(0,A)].
\end{equation}
It measures the expected logarithmic growth of a rare mutant relative to the resident. 

For \(\varepsilon\in(0,1)\), define the local exit time by
\begin{equation}\label{eq:exit_time}
\tau_\varepsilon:=\inf\{t\ge0:q_t\ge\varepsilon\},
\end{equation}
with \(\inf\varnothing=\infty\). The event \(\{\tau_\varepsilon=\infty\}\), called \emph{nonexit}, means that mutant frequency remains below \(\varepsilon\) at every generation.

For the fixed pair \(x,y\), we say that the resident fixation state \(\hat q=0\) is \emph{locally asymptotically stable in probability} if, for every sufficiently small \(\varepsilon>0\) and every \(\eta\in(0,1)\), there is a \(\delta\in(0,\varepsilon)\) such that \(0<q_0<\delta\) implies
\begin{equation}\label{eq:LAS_probability_definition}
\mathbb P_{q_0}\!\left(\tau_\varepsilon=\infty,\ \lim_{t\to\infty}q_t=0\right)\ge1-\eta.
\end{equation}
Here \(\mathbb P_{q_0}\) denotes the law of the process with initial mutant frequency \(q_0\). In what follows, we write \(\E_{q_0}\) for expectation under this law. Definition \eqref{eq:LAS_probability_definition} requires two properties to hold together with probability that can be made arbitrarily close to $1$ by choosing \(q_0\) sufficiently small. Omitting the convergence requirement in Eq. \eqref{eq:LAS_probability_definition} gives local stability in probability. A resident \(x^\ast\) is \emph{stochastically evolutionarily stable}, abbreviated SES, if the resident fixation state is locally asymptotically stable in probability against every distinct mutant \(y\in\Delta_{n-1}\). The admissible initial frequency, neighborhood, and constants in the estimates may depend on the mutant.

The following result relates the sign of \(Q(y,x)\) to local stability and gives the associated extinction rate or exit-time estimate.

\begin{result}\label{thm:Q_controls_q0}
Suppose that Assumptions \textbf{($\mathcal A1$)}--\textbf{($\mathcal A3$)} hold, and fix distinct resident and mutant phenotypes \(x,y\in\Delta_{n-1}\).
\begin{enumerate}
\item[(i)] If \(Q(y,x)<0\), then the state $\hat q=0$ is locally asymptotically stable in probability against \(y\). Moreover, for every sufficiently small \(\varepsilon\in(0,1)\) and every \(0<q_0<\varepsilon\), we have
\begin{equation}\label{eq:negative_Q_exact_rate}
\lim_{t\to\infty}\frac{\log q_t}{t}=Q(y,x)
\end{equation}
a.s. under \(\mathbb P_{q_0}\) on \(\{\tau_\varepsilon=\infty\}\).
\item[(ii)] If $Q(y,x)>0$, then for every sufficiently small $\varepsilon\in(0,1)$ and every $0<q_0<\varepsilon$, there exists $\lambda>0$ such that
\begin{equation}\label{eq:positive_Q_exit}
\E_{q_0}\!\left[e^{\lambda\tau_\varepsilon}\right]<\infty.
\end{equation}
In particular, \(\mathbb P_{q_0}(\tau_\varepsilon<\infty)=1\), and the state $\hat q=0$ is not locally stable in probability against \(y\).
\end{enumerate}
\end{result}

This stability criterion was established in \cite{ZhengLiLessardTao2017,FengMatrix2022} by combining the strong law of large numbers with Egorov's theorem. Their logarithmic decompositions [Eq.~(A2) of \cite{ZhengLiLessardTao2017} and Eq.~(7) of \cite{FengMatrix2022}] also imply the exponential extinction rate on trajectories for which \(q_t\to0\), although this rate is not stated there. Our proof in Appendix~\ref{app:local_invasion} uses exponential estimates and stopped supermartingales instead, and gives quantitative information that the earlier argument does not provide. When \(Q(y,x)<0\), it gives the explicit bound \eqref{eq:exit_probability_negative_Q} on the probability of local escape. When \(Q(y,x)>0\), it shows that escape occurs almost surely, not only that convergence to fixation has probability zero, and that the exit time has the exponential moment \eqref{eq:positive_Q_exit}.

Local escape means that the mutant reaches a small threshold frequency, it does not imply eventual mutant fixation. When \(Q(y,x)=0\), Result~\ref{thm:Q_controls_q0} gives no stability conclusion. We classify this case below for SNE residents.

For a resident \(x\), define the best-reply set as
\begin{equation}\label{eq:BR_def}
\BR(x):=\arg\max_{y\in\Delta_{n-1}}F(y,x).
\end{equation}
Following \cite{LiNoiseSNE2025}, a resident \(x^\ast\) is a \emph{stochastic Nash equilibrium}, abbreviated SNE, if
\begin{equation}\label{eq:SNE}
F(y,x^\ast)\le F(x^\ast,x^\ast),
\end{equation}
for every $y\in\Delta_{n-1}$. This is equivalent to \(Q(y,x^\ast)\le0\) for every mutant \(y\).

Note that every SES must be an SNE. Otherwise, some mutant has a positive invasion exponent and escapes by Result~\ref{thm:Q_controls_q0}\textup{(ii)}. Now, consider an SNE \(x^\ast\) and a distinct mutant \(y\). If \(y\notin\BR(x^\ast)\), then \(Q(y,x^\ast)<0\), and the mutant is resisted by Result~\ref{thm:Q_controls_q0}\textup{(i)}. It remains to examine the case \(y\in\BR(x^\ast)\setminus\{x^\ast\}\), for which \(Q(y,x^\ast)=0\). Such a mutant is called a \emph{tied best reply}.

The next result establishes the existence of at least one SNE and characterizes its best replies.

\begin{result}\label{prop:SNE_existence}\label{lem:tie_rigidity}
Under Assumptions \textup{($\mathcal A1$)}-\textup{($\mathcal A2$)}, at least one SNE exists. In addition, for any SNE \(x^\ast\), we have \(y\in\BR(x^\ast)\) if and only if
\begin{equation}\label{eq:tie_rigidity}
y^{\T}Ax^\ast=(x^\ast)^{\T}Ax^\ast\quad\text{a.s.}
\end{equation}
\end{result}

In a deterministic matrix game, tied best replies to a Nash equilibrium \(x^\ast\) satisfy \(y^{\T}Ax^\ast=(x^\ast)^{\T}Ax^\ast\) \cite{HofbauerSigmund1998}. Result~\ref{prop:SNE_existence} gives the corresponding identity for random fitness matrices, with equality holding a.s. Although an SNE is defined through expected logarithmic fitness, its tied best replies still have the same realized fitness against the resident in almost every environment. This conclusion follows from the strict concavity of the logarithm. If \(y\) and \(x^\ast\) had different realized fitnesses against $x^\ast$ with positive probability, then, for any \(0<\lambda<1\), the strategy \(y_\lambda=(1-\lambda)x^\ast+\lambda y\) would satisfy
\[
F(y_\lambda,x^\ast)>
(1-\lambda)F(x^\ast,x^\ast)+\lambda F(y,x^\ast)
=F(x^\ast,x^\ast).
\]
The last equality holds because \(y\) and \(x^\ast\) are both best replies. This contradicts the SNE condition. We call this property \emph{tie rigidity}. The existence and tie rigidity proofs are given in Appendix~\ref{app:SNE_structure}.

An SNE \(x^\ast\) is \emph{strong} if \(\BR(x^\ast)=\{x^\ast\}\). In other words, the resident has strictly greater expected logarithmic fitness against itself than any distinct mutant has against it. Equivalently, \(Q(y,x^\ast)<0\) for every \(y\ne x^\ast\). Result~\ref{thm:Q_controls_q0}\textup{(i)} shows that every strong SNE is SES, with exponential mutant extinction a.s. on nonexit.

An SNE is \emph{fully weak} if \(\BR(x^\ast)=\Delta_{n-1}\). Then every mutant is a tied best reply, so \(Q(y,x^\ast)=0\) for every \(y\). By tie rigidity, all strategies have the same realized fitness against the resident a.s. However, their fitnesses against mutants may differ, so this indifference does not imply neutral competition. A fully weak SNE can still be SES, but its stability must be determined through the frequency-dependent tests developed in Section~\ref{sec:degenerate_stability}.

More generally, an SNE that is not strong has at least one distinct tied best reply, but need not be fully weak. In that case, mutants outside \(\BR(x^\ast)\) are resisted by Result \ref{thm:Q_controls_q0}(i), while tied mutants require further analysis.
The next result characterizes strong and fully weak SNEs and shows that, when \(n=2\), every SNE belongs to one of these two classes in agreement with \cite{LiNoiseSNE2025}.

\begin{result}\label{res:strong_fully_weak}
Under Assumptions \textup{($\mathcal A1$)}-\textup{($\mathcal A2$)}, let \(x^\ast\) be an SNE.
\begin{enumerate}
\item[(i)] The equilibrium \(x^\ast\) is strong if and only if
\[
\mathbb P\!\left((y-x^\ast)^{\T}A(t)x^\ast\ne0\right)>0
\]
for every $y\in\Delta_{n-1}\setminus\{x^\ast\}$.
\item[(ii)] The equilibrium \(x^\ast\) is fully weak if and only if
\[
A(t)x^\ast=c(t)\1\quad\text{a.s.},
\]
where $c(t):=(x^\ast)^{\T}A(t)x^\ast$.
\end{enumerate}
When \(n=2\), every SNE is either strong or fully weak.
\end{result}

The proof is given in Appendix~\ref{app:SNE_structure}. The next section examines how interactions with mutants affect relative fitness, including the case in which the mean first-order effect vanishes.

\section{Stability and time scales for tied mutants}\label{sec:degenerate_stability}

Now, let \(x^\ast\) be an SNE and fix a distinct tied mutant \(y\in\BR(x^\ast)\setminus\{x^\ast\}\). Its invasion exponent is zero, so stability depends on how relative fitness changes with mutant frequency.

Let \(c(t):=(x^\ast)^{\T}A(t)x^\ast\) denote the resident fitness at fixation, and define the normalized fitnesses
\[
\alpha(t):=\frac{y^{\T}A(t)y}{c(t)}
\quad\text{and}\quad
\beta(t):=\frac{(x^\ast)^{\T}A(t)y}{c(t)}.
\]
Both lie in \([m/M,M/m]\) by \textbf{($\mathcal A2$)}. Tie rigidity gives \(y^{\T}A(t)x^\ast=c(t)\) a.s., so the fitness difference between the two phenotypes comes entirely from interactions with mutants. Substituting into \eqref{eq:odds_recursion} gives
\begin{equation}\label{eq:zeroH_odds}
\begin{aligned}
z_{t+1}&=z_t\frac{(1-q_t)y^{\T}A(t)x^\ast+q_ty^{\T}A(t)y}{(1-q_t)(x^\ast)^{\T}A(t)x^\ast+q_t(x^\ast)^{\T}A(t)y}\\
&=z_t\frac{1+\alpha(t)z_t}{1+\beta(t)z_t}.
\end{aligned}
\end{equation}

Near resident fixation, where \(0<z_t\ll1\), the logarithmic increment satisfies
\begin{equation}\label{eq:tied_log_expansion}
\begin{aligned}
&\log\frac{z_{t+1}}{z_t}\\
=&\log(1+\alpha(t)z_t)-\log(1+\beta(t)z_t)\\
=&(\alpha(t)-\beta(t))z_t-\frac12\bigl(\alpha(t)^2-\beta(t)^2\bigr)z_t^2
+O(z_t^3).
\end{aligned}
\end{equation}
The fitness bounds make the remainder uniform over environmental realizations. Let \(\mathcal F_t:=\sigma(A(0),\ldots,A(t-1))\) for \(t\ge1\) represent the information generated by the environmental history before generation \(t\), with \(\mathcal F_0\) trivial. Since \(q_t\) is \(\mathcal F_t\)-measurable and \(A(t)\) is independent of \(\mathcal F_t\), taking the conditional expectation of \eqref{eq:tied_log_expansion} gives
\[
\E_{q_0}\!\left[\log\frac{z_{t+1}}{z_t}\,\middle|\,\mathcal F_t\right]
=-H(y,x^\ast)z_t+O(z_t^2),
\]
where
\begin{equation}\label{eq:H_def}
H(y,x^\ast):=\E\left[\beta(t)-\alpha(t)\right].
\end{equation}
The coefficient \(H(y,x^\ast)\) measures the resident's expected fitness advantage in interactions with mutants, normalized by resident fitness at fixation. A positive value opposes invasion, while a negative value favors mutant growth.

\begin{result}\label{thm:SES_sufficient}
Let \(x^\ast\) be an SNE and \(y\in\BR(x^\ast)\setminus\{x^\ast\}\). Then the following conclusions hold.
\begin{enumerate}
\item[(i)] If \(H(y,x^\ast)>0\), then the state \(\hat q=0\) is locally asymptotically stable in probability against \(y\). Moreover, for every sufficiently small \(\varepsilon\in(0,1)\) and every \(0<q_0<\varepsilon\), we have
\begin{equation}\label{eq:tied_inverse_time_rate}
\lim_{t\to\infty}tq_t=\frac{1}{H(y,x^\ast)}
\end{equation}
a.s. under \(\mathbb P_{q_0}\) on \(\{\tau_\varepsilon=\infty\}\).
\item[(ii)] If \(H(y,x^\ast)<0\), then, for every sufficiently small \(\varepsilon\in(0,1)\) and every \(0<q_0<\varepsilon\),
\begin{equation}\label{eq:tied_negative_H_escape}
\E_{q_0}[\tau_\varepsilon]<\infty.
\end{equation}
In particular, \(\mathbb P_{q_0}(\tau_\varepsilon<\infty)=1\), and the state \(\hat q=0\) is not locally stable in probability against \(y\).
\end{enumerate}
\end{result}

The stability criteria based on the sign of \(H(y,x^\ast)\) were established for two pure strategies in \cite{ZhengLiLessardTao2017,FengMatrix2022}, using reciprocal-odds arguments together with the strong law of large numbers and Egorov's theorem. Li \textit{et al.}\ \cite{LiNoiseSNE2025} applied them to weak SNEs, and Zheng, Tao, and Li \cite{ZhengTaoLi2025} extended them to mixed strategies with three pure strategies through a small-frequency approximation. These decompositions [Eq.~(A19) of \cite{ZhengLiLessardTao2017}, Eq.~(15) of \cite{FengMatrix2022}, and Eq.~(A5) of \cite{ZhengTaoLi2025}] also suggest the inverse-time extinction rate, although it is not stated as a result there. Result~\ref{thm:SES_sufficient} applies to every tied best reply of an SNE, for any finite number of pure strategies, and states this rate explicitly in \eqref{eq:tied_inverse_time_rate}. Its proof, given in Appendix~\ref{app:tied_first_order}, uses stopped drift estimates on the exact recursion. When \(H(y,x^\ast)>0\), it also gives the explicit bound \eqref{eq:tied_exit_probability} on the probability of local escape. When \(H(y,x^\ast)<0\), it shows that the expected exit time is finite, so escape occurs almost surely.

Now suppose that \(H(y,x^\ast)=0\). The linear term in \eqref{eq:tied_log_expansion} has zero expectation but may still fluctuate across environments. To examine this case, define
\begin{subequations}
\begin{align}
K(y,x^\ast)&:=\frac12\E[\alpha^2-\beta^2],
\label{eq:zeroH_K}\\
V(y,x^\ast)&:=\E[(\alpha-\beta)^2],
\label{eq:zeroH_V}
\end{align}
\end{subequations}
where the time index is omitted inside expectations.

Using \eqref{eq:tied_log_expansion}, the conditional mean and variance of logarithmic growth satisfy
\begin{subequations}\label{eq:zeroH_log_moments}
\begin{align}
\E_{q_0}\!\left[\log\frac{z_{t+1}}{z_t}\,\middle|\,\mathcal F_t\right]
&=-K(y,x^\ast)z_t^2+O(z_t^3),\\
\operatorname{Var}_{q_0}\!\left(\log\frac{z_{t+1}}{z_t}\,\middle|\,\mathcal F_t\right)
&=V(y,x^\ast)z_t^2+O(z_t^3).
\end{align}
\end{subequations}
The coefficient \(K(y,x^\ast)\) determines the quadratic mean term, while \(V(y,x^\ast)\) measures the leading fluctuations. Since \(0=H(y,x^\ast)=\E[\beta-\alpha]\), we have \(\E[\alpha]=\E[\beta]\), and hence
\[
K(y,x^\ast)
=\frac12\bigl(\operatorname{Var}(\alpha)-\operatorname{Var}(\beta)\bigr).
\]
Thus \(K(y,x^\ast)>0\) means that the normalized mutant fitness against mutants has greater variance than the corresponding resident fitness. This variance difference gives a negative quadratic contribution to expected logarithmic relative growth.

If \(V(y,x^\ast)=0\), then \(\alpha=\beta\) a.s., so \(H(y,x^\ast)=K(y,x^\ast)=0\). Equation~\eqref{eq:zeroH_odds} gives \(z_{t+1}=z_t\) a.s. for every \(t\ge0\). These equalities hold simultaneously a.s., and induction gives \(q_t=q_0\) for every \(t\ge0\). Thus, if \(q_0<\varepsilon\), the mutant frequency remains below \(\varepsilon\), whereas convergence to zero is impossible from any \(q_0>0\).

\begin{result}\label{res:exact_neutrality}
Let \(x^\ast\) be an SNE and \(y\in\BR(x^\ast)\setminus\{x^\ast\}\). If \(V(y,x^\ast)=0\), then \(H(y,x^\ast)=K(y,x^\ast)=0\) and, for every \(q_0\in(0,1)\),
\begin{equation}\label{eq:zeroH_exact_neutrality}
\mathbb P_{q_0}\!\left(q_t=q_0\text{ for every }t\ge0\right)=1.
\end{equation}
The state $\hat q=0$ is stable in probability but is not locally asymptotically stable in probability against \(y\).
\end{result}

We consider the remaining tied mutants, for which \(H(y,x^\ast)=0\) and \(V(y,x^\ast)>0\), including the case \(K(y,x^\ast)=0\). By \eqref{eq:zeroH_log_moments}, the conditional mean and the conditional variance of the log-increments of $z_t$ are both of order $z_t^2$. Near $z_t=0$, drift and noise therefore vanish at the same rate, neither dominates, and first-order arguments give no conclusion.

To study the fluctuations that persist as the mutant becomes rare, we use the reciprocal odds
\[
w_t:=\frac{1}{z_t}=\frac{1-q_t}{q_t}.
\]
Mutant extinction corresponds to \(w_t\to\infty\), while exit from \(q_t<\varepsilon\) occurs when \(w_t\) first falls to or below \(w_\varepsilon:=(1-\varepsilon)/\varepsilon\). Rewriting \eqref{eq:zeroH_odds} in terms of the reciprocals gives
\begin{equation}\label{eq:zeroH_reciprocal_increment}
w_{t+1}-w_t=\phi(w_t,A(t)),
\end{equation}
where $\phi(w,A):=\dfrac{w(\beta-\alpha)}{w+\alpha}$.
The increments are uniformly bounded since
\[
|\phi(w,A)|\le|\beta-\alpha|\le\frac{M-m}{m}
\quad\text{for}\;\;w>0.
\]
In addition, the first two moments satisfy, as \(w\to\infty\),
\begin{subequations}\label{eq:zeroH_increment_moments}
\begin{align}
\nu_1(w)&:=\E[\phi(w,A)]=\frac{K(y,x^\ast)+V(y,x^\ast)/2}{w}+O(w^{-2}),\\
\nu_2(w)&:=\E[\phi(w,A)^2]=V(y,x^\ast)+O(w^{-1}).
\end{align}
\end{subequations}
Unlike in the $z_t$ coordinates, the reciprocal transformation removes the boundary degeneracy. As $w\to\infty$, its increments remain uniformly bounded with a non-vanishing asymptotic variance $V(y,x^\ast)>0$, while its mean drift decays at rate $O(w^{-1})$. Consequently, the process satisfies the non-degeneracy conditions of Lamperti's theory \cite{Lamperti1960,Lamperti1963,DenisovKorshunovWachtel2025}. More precisely, the theory compares the directional push $2w\nu_1(w)$ against the diffusive spread $\nu_2(w)$. The sign of
\[
2w\nu_1(w)-\nu_2(w)=2K(y,x^\ast)+O(w^{-1})
\]
determines whether $w_t\to\infty$ on nonexit, an event of probability close to one for small initial mutant frequency, which means attraction to resident fixation ($K(y,x^\ast)>0$), or $w_t$ falls to or below $w_\varepsilon$ almost surely, which means local escape of the mutant ($K(y,x^\ast)<0$). At $K(y,x^\ast)=0$ the leading term vanishes, but the $O(w^{-1})$ correction is too small to prevent escape, so escape still occurs almost surely. When escape occurs, the sign of
\[
2w\nu_1(w)+\nu_2(w)=2\bigl(K(y,x^\ast)+V(y,x^\ast)\bigr)+O(w^{-1})
\]
determines whether the expected exit time is finite ($K(y,x^\ast)<-V(y,x^\ast)$) or infinite ($-V(y,x^\ast)<K(y,x^\ast)\le 0$). At $K(y,x^\ast)=-V(y,x^\ast)$ this quantity is $O(w^{-1})$ and the sign gives no conclusion; a separate argument shows that the expected exit time is still infinite. We summarize these findings in the following two results.
\begin{result}
\label{thm:zeroH_positiveK}
Let \(x^\ast\) be an SNE and \(y\in\BR(x^\ast)\setminus\{x^\ast\}\). Suppose that \(H(y,x^\ast)=0\) and \(K(y,x^\ast)>0\). The following conclusions hold.
\begin{enumerate}
\item[(i)] The state \(\hat q=0\) is locally asymptotically stable in probability against \(y\).
\item[(ii)] For every sufficiently small \(\varepsilon\in(0,1)\) and every \(0<q_0<\varepsilon\),
\begin{equation}\label{eq:zeroH_logarithmic_rate}
\lim_{t\to\infty}\frac{\log q_t}{\log t}=-\frac12
\end{equation}
a.s. under \(\mathbb P_{q_0}\) on \(\{\tau_\varepsilon=\infty\}\).
\end{enumerate}
\end{result}

Appendix~\ref{app:zeroH_stability} proves local attraction and the exit-probability bound in part~\textup{(i)}, while Appendix~\ref{app:zeroH_rates} proves the logarithmic extinction rate in part~\textup{(ii)}.

\begin{result}\label{thm:zeroH_escape}
Let \(x^\ast\) be an SNE and \(y\in\BR(x^\ast)\setminus\{x^\ast\}\). Suppose that \(H(y,x^\ast)=0\), \(K(y,x^\ast)\le0\), and \(V(y,x^\ast)>0\). For every sufficiently small \(\varepsilon\in(0,1)\) and every \(0<q_0<\varepsilon\), the following conclusions hold.
\begin{enumerate}
\item[(i)] We have
\begin{equation}\label{eq:zeroH_escape}
\mathbb P_{q_0}(\tau_\varepsilon<\infty)=1.
\end{equation}
In particular, the state \(\hat q=0\) is not locally stable in probability against \(y\).
\item[(ii)] The expected exit time satisfies
\begin{subequations}\label{eq:zeroH_mean_exit}
\begin{align}
\E_{q_0}[\tau_\varepsilon]&<\infty
&&\text{if }K(y,x^\ast)<-V(y,x^\ast),\\
\E_{q_0}[\tau_\varepsilon]&=\infty
&&\text{if }-V(y,x^\ast)\le K(y,x^\ast)\le0.
\end{align}
\end{subequations}
\end{enumerate}
\end{result}
The proof of Result~\ref{thm:zeroH_escape}\textup{(i)} is given in Appendix~\ref{app:zeroH_stability}. The proof of part~\textup{(ii)}, including the critical case \(K(y,x^\ast)=-V(y,x^\ast)\), is given in Appendix~\ref{appendix_proof_thm:zeroH_mean_exit}.

Throughout the range \(-V(y,x^\ast)\le K(y,x^\ast)\le 0\), exit is certain but its expected time is infinite. In the stable regimes, the unconditional expected exit time is also infinite for sufficiently small introductions, because there is a positive probability of never leaving the neighborhood. Then, an infinite expected exit time alone does not distinguish stability from instability.

Note that local escape a.s. also rules out convergence to the resident fixation state. To see this, fix \(\varepsilon>0\) such that
\[
\mathbb P_{q_0}(\tau_\varepsilon<\infty)=1
\quad\text{for every }q_0\in(0,\varepsilon).
\]
For each integer \(T\ge0\), define 
$$E_T:=\bigcap_{s=T}^{\infty}\{q_s<\varepsilon\}.$$ Thus \(E_T\) is the event that mutant frequency remains below \(\varepsilon\) from generation \(T\) onward. By independence of the future environments from \(\mathcal F_T\), the Markov property gives
\[
\mathbb P_{q_0}(E_T\mid\mathcal F_T)=\mathbf1_{\{q_T<\varepsilon\}}\mathbb P_{q_T}(\tau_\varepsilon=\infty)=0\quad\text{a.s.}
\]
Taking expectations yields \(\mathbb P_{q_0}(E_T)=0\). On the other hand, convergence to zero requires mutant frequency to remain below \(\varepsilon\) after some finite generation, so
\[
\left\{\lim_{t\to\infty}q_t=0\right\}
\subseteq\bigcup_{T=0}^{\infty}E_T.
\]
Consequently, we obtain
\[
\mathbb P_{q_0}\!\left(\lim_{t\to\infty}q_t=0\right)\le\sum_{T=0}^{\infty}\mathbb P_{q_0}(E_T)=0
\]
for every $q_0\in(0,1)$. This conclusion applies to the escaping regimes in Results~\ref{thm:Q_controls_q0}\textup{(ii)}, \ref{thm:SES_sufficient}\textup{(ii)}, and~\ref{thm:zeroH_escape}\textup{(i)}, regardless of whether the expected exit time is finite or not. These regimes therefore also fail the convergence requirement.

\begin{figure*}[t]
\centering
\includegraphics[width=\textwidth]{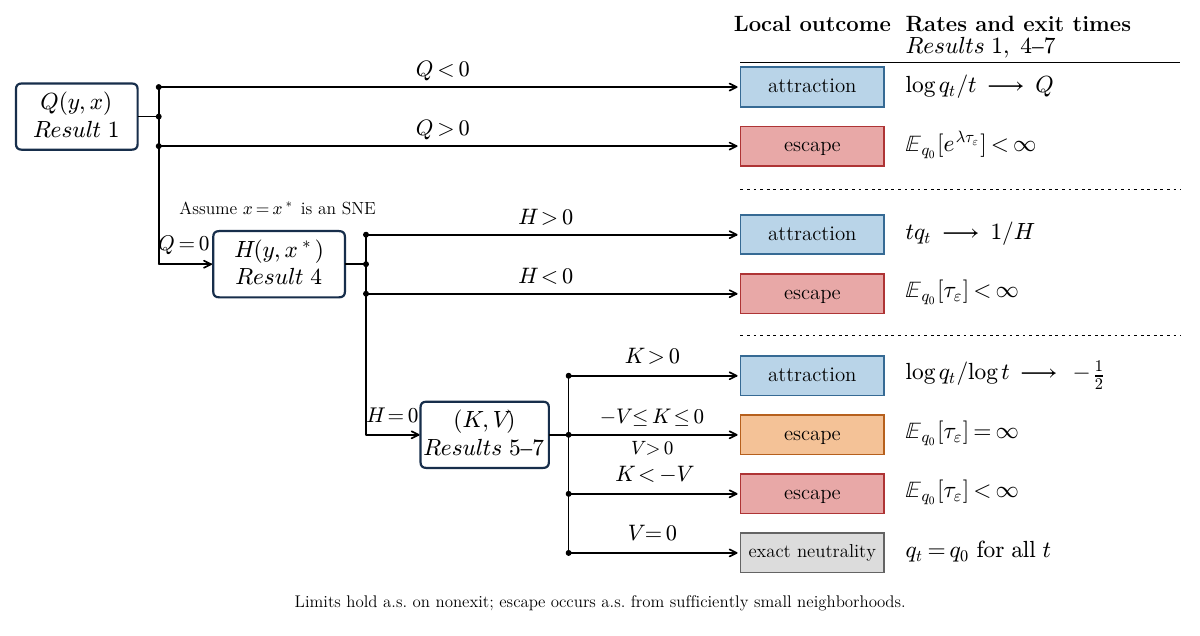}
\caption{Local invasion outcomes and their time scales under \textbf{($\mathcal A1$)}--\textbf{($\mathcal A3$)}. The test based on \(Q(y,x)\) applies to any fixed resident--mutant pair. The continuation through \(H(y,x^\ast)\), \(K(y,x^\ast)\), and \(V(y,x^\ast)\) assumes an SNE resident and a distinct tied mutant. Attraction denotes local asymptotic stability in probability. The extinction limits hold a.s. on \(\{\tau_\varepsilon=\infty\}\), while escape occurs a.s. from sufficiently small neighborhoods. The neighborhood may depend on the mutant. For \(Q(y,x)>0\), the exponential moment is finite for some \(\lambda>0\). The orange branch represents a.s. escape with an infinite expected exit time. When \(V(y,x^\ast)=0\), mutant frequency remains constant a.s.}
\label{fig:invasion_classification}
\end{figure*}

Figure~\ref{fig:invasion_classification} summarizes the local invasion outcomes, extinction rates, and expected exit times.

The next section combines these results into a criterion for SES and expresses it in terms of feasible mutant directions.

\section{Characterization of stochastic evolutionary stability}\label{sec:SES_classification}
The preceding stability results give a necessary and sufficient condition for an SNE to be SES. Mutants outside \(\BR(x^\ast)\) are resisted by Result~\ref{thm:Q_controls_q0}\textup{(i)}. Distinct tied best replies are resisted when \(H(y,x^\ast)>0\), or when \(H(y,x^\ast)=0\) and \(K(y,x^\ast)>0\), by Results~\ref{thm:SES_sufficient}\textup{(i)} and~\ref{thm:zeroH_positiveK}\textup{(i)}. All remaining tied mutants either escape locally a.s. or compete neutrally with the resident, so the attraction requirement fails. These conclusions give the following characterization.

\begin{result}\label{cor:zeroH_SES}
Under Assumptions \textbf{($\mathcal A1$)}--\textbf{($\mathcal A3$)}, an SNE \(x^\ast\) is SES if and only if every \(y\in\BR(x^\ast)\setminus\{x^\ast\}\) satisfies
\begin{equation}\label{eq:zeroH_exact_SES}
H(y,x^\ast)>0
\quad\text{or}\quad
\bigl[H(y,x^\ast)=0\ \text{and}\ K(y,x^\ast)>0\bigr].
\end{equation}
\end{result}

A strong SNE satisfies this condition because it has no distinct tied best replies.

We can also express this criterion in terms of mutant displacements to show how the resident's support affects stability. At a boundary resident, a feasible displacement must have nonnegative components for actions absent from the resident. Define the feasible neutral directions by
\[
\mathcal N(x^\ast):=
\left\{
h\in\mathbb R^n:
\begin{array}{l}
\1^{\T}h=0,\\
h_i\ge0\text{ whenever }x_i^\ast=0,\\
h^{\T}A(t)x^\ast=0\text{ a.s.}
\end{array}
\right\}.
\]
By tie rigidity, a strategy \(y\in\Delta_{n-1}\) is a distinct tied best reply to \(x^\ast\) if and only if \(y-x^\ast\in\mathcal N(x^\ast)\setminus\{0\}\). Thus each nonzero \(h\in\mathcal N(x^\ast)\) generates tied mutants \(y_s=x^\ast+sh\) for every \(s>0\) such that \(y_s\in\Delta_{n-1}\).
Thus, for every nonzero \(h\in\mathcal N(x^\ast)\), the strategy \(y_s=x^\ast+sh\) is a distinct tied best reply whenever it belongs to the simplex, as it does for all sufficiently small \(s>0\). With \(c(t)=(x^\ast)^{\T}A(t)x^\ast\), define
\[
B_h:=\frac{(x^\ast)^{\T}A(t)h}{c(t)},
\qquad
C_h:=\frac{h^{\T}A(t)h}{c(t)}.
\]
The coefficients satisfy the exact identities
\begin{equation}\label{eq:directional_HK_scaling}
\begin{aligned}
H(y_s,x^\ast)&=-s^2\E[C_h],\\
K(y_s,x^\ast)&=s^3\E[B_hC_h]+\frac{s^4}{2}\E[C_h^2],
\end{aligned}
\end{equation}
for every feasible \(s>0\), where the second identity requires \(\E[C_h]=0\). Here \(s\) measures displacement in strategy space and is distinct from the mutant population frequency \(q_t\). Together with Result~\ref{cor:zeroH_SES}, these identities yield the following directional criterion.

\begin{result}
\label{prop:SES_directional_refinement}
Under Assumptions \textbf{($\mathcal A1$)}--\textbf{($\mathcal A3$)}, an SNE \(x^\ast\) is SES if and only if every nonzero \(h\in\mathcal N(x^\ast)\) satisfies
\begin{equation}\label{eq:SES_directional_refinement}
\begin{aligned}
&\E[C_h]<0,\\
&\text{or}\quad
\E[C_h]=0,\quad
\E[B_hC_h]\ge0,\quad
\E[C_h^2]>0.
\end{aligned}
\end{equation}
\end{result}
The proof of this criterion is given in Appendix~\ref{app:SES_consequences}.

At an interior resident, both \(h\) and \(-h\) are feasible neutral directions. Since \(B_{-h}=-B_h\) and \(C_{-h}=C_h\), applying the directional criterion to both directions gives \(\E[B_hC_h]\ge0\) and \(-\E[B_hC_h]\ge0\) whenever \(\E[C_h]=0\). Hence \(\E[B_hC_h]=0\), which also satisfies the inequality in the general criterion. At a boundary resident, the opposite direction may be infeasible, allowing \(\E[B_hC_h]>0\). 

The directional criterion also shows that resistance to every distinct mutant in a neighborhood of an SNE is equivalent to resistance to every distinct mutant in the simplex. Indeed, testing mutants \(x^\ast+sh\) for arbitrarily small \(s>0\) forces the conditions in \eqref{eq:SES_directional_refinement} along every feasible neutral direction. The identities in \eqref{eq:directional_HK_scaling} then ensure resistance along the entire feasible ray, while mutants outside the best-reply set are already resisted. The converse is immediate. This equivalence concerns proximity in strategy space; the admissible initial mutant frequency may still depend on the mutant.

\section{Examples}\label{sec:examples}

In all three examples, the resident is a fully weak SNE and \(H(y,x^\ast)=0\) in the cases of interest. The earlier criteria, which require a strict sign of \(H\) \cite{ZhengLiLessardTao2017,FengMatrix2022,ZhengTaoLi2025}, therefore give no conclusion, and stability is decided by \(K\) and \(V\). In the first two examples \(\beta(t)=1\), so \(K=-H+V/2\), and \(H=0\) gives \(K=V/2>0\) whenever \(V>0\). They differ in how \(H=0\) arises. In the cyclic game, \(H=0\) holds only at a threshold of the noise level, below which fluctuations remove the resistance of an ESS of the mean game. In the neutral mean game, \(H=0\) holds for every mutant and every noise level, and fluctuations create resistance that the mean game lacks. In the third example, \(\beta(t)\) varies and \(K\) changes sign, so mutants close to the resident escape although both pure mutants are resisted. Conclusions concern fixed mutants near resident fixation, and extinction rates hold a.s. on nonexit.

\subsection{Stability at the threshold in a cyclic game}
\label{ex:rps}

We revisit the cyclic family of \cite[Example~1]{ZhengTaoLi2025} at the threshold where \(H\) vanishes. Let \(\varsigma_t\) be independent and take the values \(d_0\pm\sigma\) with equal probabilities, and set
\begin{equation}\label{eq:rps-matrix}
A(t)=\1\1^{\T}+
\begin{pmatrix}
0&-b_0&b_0+\varsigma_t\\
b_0+\varsigma_t&0&-b_0\\
-b_0&b_0+\varsigma_t&0
\end{pmatrix},
\end{equation}
with \(0<b_0<1\), \(\sigma\ge0\), and \(d_0-\sigma>-b_0\), so that all entries are positive and bounded. (After exchanging the pure strategies \(2\) and \(3\), this is Eq.~(20) of \cite{ZhengTaoLi2025} with \(a(t)=1\), \(b(t)=1+b_0+\varsigma_t\), and \(c(t)=1-b_0\); their condition \(\bar b>\bar a>\bar c\) holds since \(b_0+d_0>0\), and their condition (22) reads \(\E[\varsigma_t/(3+\varsigma_t)]>0\)).

Every row and column of \(A(t)\) sums to \(3+\varsigma_t\). Hence, with \(x^\ast=\1/3\) and \(c(t)=1+\varsigma_t/3\), we have \(A(t)x^\ast=c(t)\1\) and \((x^\ast)^{\T}A(t)=c(t)\1^{\T}\). Thus \(x^\ast\) is a fully weak SNE by Result~\ref{res:strong_fully_weak}\textup{(ii)}, and \(\beta(t)=1\) for every mutant.

For \(y\ne x^\ast\), let \(d:=\|y-x^\ast\|^2\), so that \(\|y\|^2=1/3+d\). Each product \(y_iy_j\) with \(i\ne j\) appears in \(y^{\T}A(t)y\) with the coefficient \(\varsigma_t\) from the second matrix. Hence
\[
y^{\T}A(t)y=1+\frac{\varsigma_t}{2}\bigl(1-\|y\|^2\bigr)=c(t)-\frac{d\varsigma_t}{2},
\]
so that
\[
\alpha(t)=1-\frac{d\varsigma_t}{2c(t)}.
\]
Averaging over \(\varsigma_t=d_0\pm\sigma\) gives
\[
H(y,x^\ast)=\frac d2\,\E\!\left[\frac{\varsigma_t}{c(t)}\right]
=\frac{3d}{2}\,\frac{d_0(3+d_0)-\sigma^2}{(3+d_0)^2-\sigma^2}.
\]
Since \(d_0-\sigma>-b_0>-1\), the denominator \((3+d_0-\sigma)(3+d_0+\sigma)\) is positive and \(d_0>-1\). Hence \(H(y,x^\ast)\) has the sign of \(d_0-d_0^\ast(\sigma)\), where \(d_0^\ast(\sigma):=(\sqrt{9+4\sigma^2}-3)/2\) is the larger root of \(d_0(3+d_0)=\sigma^2\). At \(d_0=d_0^\ast(\sigma)\) with \(\sigma>0\), we have \(V(y,x^\ast)>0\) and hence \(K(y,x^\ast)=V(y,x^\ast)/2>0\). Result~\ref{cor:zeroH_SES} therefore gives, for \(\sigma>0\),
\begin{equation}\label{eq:rps-threshold}
x^\ast\text{ is SES}
\quad\Longleftrightarrow\quad
d_0\ge d_0^\ast(\sigma).
\end{equation}
The strict condition of \cite[Eq.~(22)]{ZhengTaoLi2025} corresponds to \(d_0>d_0^\ast(\sigma)\), so it is sufficient but not necessary.

The threshold also changes the extinction rate. Above it, \(tq_t\to1/H(y,x^\ast)\) by Result~\ref{thm:SES_sufficient}\textup{(i)}; at it, \(\log q_t/\log t\to-1/2\) by Result~\ref{thm:zeroH_positiveK}\textup{(ii)}, both a.s. on nonexit. Below it, the mutant escapes a.s. with finite mean exit time by Result~\ref{thm:SES_sufficient}\textup{(ii)}. For the mean matrix \(\overline A=\E[A(t)]\), the same calculation gives \((x^\ast)^{\T}\overline A y-y^{\T}\overline A y=d_0d/2\), so \(x^\ast\) is an ESS of the mean game for every \(d_0>0\). Parameters with \(0<d_0<d_0^\ast(\sigma)\) thus show that fluctuations can remove invasion resistance.

\subsection{Noise-induced stability in a neutral mean game}\label{ex:fully_weak_zero_H}

For \(n\ge2\), let \(x^\ast=\1/n\), \(P=I-\1\1^{\T}/n\), and
\begin{equation}\label{eq:fully_weak_SES_matrix}
A(t)=2\1\1^{\T}+\xi_tP,
\end{equation}
where the independent variables \(\xi_t\) take the values \(\pm r\) with equal probabilities and \(0<r<1\). The mean matrix \(2\1\1^{\T}\) is neutral, so in the mean game the mutant frequency stays constant. For \(n=2\), this is Example~A.2 of \cite{LiNoiseSNE2025} (with \(a=\hat v=1/2\), \(b=c=2\), and \(\eta_t=2\xi_t\) in their notations). Li \textit{et al.}\ showed that the noise creates a unique weak SNE. Their sufficient condition for stability corresponds to \(H(y,x^\ast)>0\), but here \(H(y,x^\ast)=0\) for every mutant [Eq.~(A.32) of \cite{LiNoiseSNE2025}], so the stability of this SNE was left open. For \(n=3\), the example also fails the strict condition (19) of \cite{ZhengTaoLi2025}, which is stated there as necessary and sufficient, so that criterion would classify \(x^\ast\) as not SES.

Since \(A(t)x^\ast=2\1\), \(x^\ast\) is a fully weak SNE, and since \((x^\ast)^{\T}P=0\), we have \((x^\ast)^{\T}A(t)y=2\), so \(\beta(t)=1\). For \(y\ne x^\ast\), we have \(Py=y-x^\ast\), and since \(P^{\T}=P=P^2\), \(y^{\T}Py=\|y-x^\ast\|^2\). Hence \(y^{\T}A(t)y=2+\xi_t\|y-x^\ast\|^2\) and \(\alpha(t)=1+\xi_t\|y-x^\ast\|^2/2\), which give $H(y,x^\ast)=0$ and
\begin{equation}\label{eq:noise_induced_SES_coefficients}
K(y,x^\ast)=\frac12V(y,x^\ast)=\frac{r^2}{8}\|y-x^\ast\|^4>0.
\end{equation}
By Result~\ref{thm:zeroH_positiveK}\textup{(i)}, \(x^\ast\) is SES for every \(n\ge2\). The resistance comes from the noise itself: the mutant's fitness against itself has mean \(2\) but positive variance, which lowers its long-run growth because growth is multiplicative.

Appendix~\ref{ex:intermediate_best_replies} shows, for \(k=n\), that \(x^\ast\) is the unique SNE and that, for each fixed \(y\ne x^\ast\), \(q_t\to0\) a.s. from every deterministic \(q_0\in(0,1)\). Combined with Result~\ref{thm:zeroH_positiveK}\textup{(ii)} and the Markov property, this gives \(\log q_t/\log t\to-1/2\) a.s. from every such \(q_0\). The appendix also extends the construction to an SES with an intermediate best-reply set.

\subsection{Resisted pure mutants, escaping mixed mutants}\label{ex:pure_mutants_not_ses}

Consider a two-strategy game with two equally likely environments, chosen independently in each generation:
\begin{equation}\label{eq:two_strategy_counterexample_matrices}
A_+=\frac18
\begin{pmatrix}
13&3\\
5&11
\end{pmatrix},
\qquad
A_-=\frac18
\begin{pmatrix}
3&13\\
11&5
\end{pmatrix}.
\end{equation}
These matrices are a specialization of the family studied by Li \textit{et al.} in \cite[Example~A.2, Eq.~(A.31)]{LiNoiseSNE2025}: in their notation, \(a=3/8\), \(\hat v=1/2\), and \(b=c=1\), with \(\eta_t=2\) giving \(A_+\) and \(\eta_t=-2\) giving \(A_-\).

Let \(x^\ast=(1/2,1/2)^{\T}\). Since \(A_\pm x^\ast=\1\), every strategy \(y\) has realized fitness \(y^{\T}A(t)x^\ast=1\) against \(x^\ast\), so \(Q(y,x^\ast)=0\) for all \(y\). Hence \(x^\ast\) is an SNE, and it is fully weak by Result~\ref{res:strong_fully_weak}\textup{(ii)}, with resident fitness \(c(t)=1\). Write each distinct mutant as \(y_s=(1/2+s,1/2-s)^{\T}\), where \(-1/2\le s\le1/2\) and \(s\ne0\). Let \(\xi_t=1\) when \(A(t)=A_+\) and \(\xi_t=-1\) when \(A(t)=A_-\). Direct calculation gives the normalized fitnesses
\[
\alpha(t)=1+\xi_t\left(\frac{s}{4}+2s^2\right),
\qquad
\beta(t)=1+\frac{\xi_t s}{4}.
\]
Averaging over the two environments gives \(H(y_s,x^\ast)=0\) and
\begin{equation}\label{eq:two_strategy_counterexample_coefficients}
\begin{aligned}
K(y_s,x^\ast)&=2s^3\left(s+\frac14\right),\\
V(y_s,x^\ast)&=4s^4.
\end{aligned}
\end{equation}

Both pure mutants, corresponding to \(s=\pm1/2\), satisfy \(K(y_s,x^\ast)>0\) and are resisted by Result~\ref{thm:zeroH_positiveK}\textup{(i)}. However, \(K(y_s,x^\ast)\le0\) and \(V(y_s,x^\ast)>0\) for \(-1/4\le s<0\). By Result~\ref{thm:zeroH_escape}\textup{(i)}, each mutant in this interval escapes locally a.s. Since the interval contains mutants arbitrarily close to \(x^\ast\), resistance to both pure mutants does not imply that \(x^\ast\) is SES.

Within the escaping interval, the mean exit time is determined by the sign of
\[
K(y_s,x^\ast)+V(y_s,x^\ast)=2s^3\left(3s+\frac14\right).
\]
For each fixed mutant and every sufficiently small \(\varepsilon>0\), Result~\ref{thm:zeroH_escape}\textup{(ii)} gives \(\E_{q_0}[\tau_\varepsilon]=\infty\) for \(-1/4\le s\le-1/12\) and \(\E_{q_0}[\tau_\varepsilon]<\infty\) for \(-1/12<s<0\), whenever \(0<q_0<\varepsilon\).

Figure~\ref{fig:example_exit_times}\textup{(a)} summarizes these regimes by plotting \(K(y_s,x^\ast)/V(y_s,x^\ast)\) against \(s\): both pure mutants lie in the attracting region, the line \(K=-V\) at \(s=-1/12\) separates finite from infinite mean exit times, and the escaping interval \(-1/4\le s<0\) reaches the resident.

\subsection{Numerical illustration of extinction rates and exit times}
\label{subsec:numerical_illustrations}

\begin{figure*}[t]
\centering
\includegraphics[width=\textwidth]{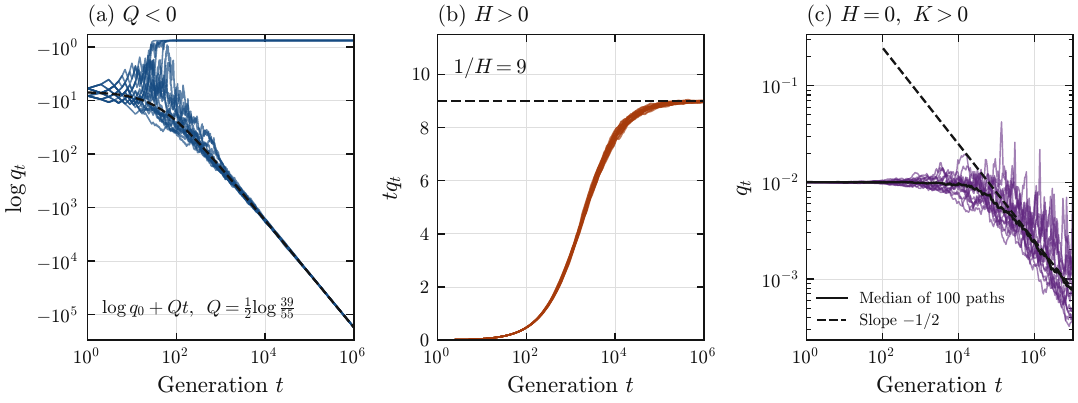}
\caption{Three extinction regimes under the exact recursion \eqref{eq:qrec}. (a) The two-strategy game \eqref{eq:two_strategy_counterexample_matrices} of Section~\ref{ex:pure_mutants_not_ses}, with resident \(e_2\), mutant \(e_1\), and \(q_0=10^{-3}\). All 20 simulated trajectories of \(\log q_t\) are shown over \(10^6\) generations, on a logarithmic time axis and a symmetric logarithmic vertical axis. The dashed curve is \(\log q_0+Q(e_1,e_2)\,t\), with \(Q(e_1,e_2)=\tfrac12\log(39/55)\). Five paths leave the neighborhood of resident fixation early and settle at \(q=1/2\), a constant interior equilibrium at which both strategies have equal fitness in both environments. This illustrates that the stability in Result~\ref{thm:Q_controls_q0}\textup{(i)} is local and holds in probability, not for every path. (b) The cyclic game of Section~\ref{ex:rps}, with \(b_0=0.9\), \(d_0=0.6\), \(\sigma=0.9\), resident \(x^\ast=\1/3\), and mutant \(e_1\). All 20 simulated trajectories of \(tq_t\) are shown, from \(q_0=0.005\) through \(10^6\) generations; the dashed line marks \(1/H(e_1,x^\ast)=9\). (c) The same game and resident--mutant pair at the threshold \(d_0=d_0^\ast(0.9)\), with \(q_0=0.01\). Of 100 simulated trajectories of \(q_t\) through \(10^7\) generations, the first 15 are shown, together with the pointwise median of all 100. The median is descriptive; the dashed line has log--log slope \(-1/2\) and specifies no asymptotic prefactor. The theoretical limits hold a.s. on nonexit from sufficiently small neighborhoods of resident fixation, whose size is not determined by these simulations.}
\label{fig:example_extinction_rates}
\end{figure*}

We simulate the exact recursion \eqref{eq:qrec} to illustrate extinction rates and exit times. Figure~\ref{fig:example_extinction_rates}\textup{(a)} uses the two-strategy game in Section~\ref{ex:pure_mutants_not_ses}, with resident \(e_2\) and mutant \(e_1\). Here
\[
Q(e_1,e_2)=\frac12\log\frac{39}{55}\approx-0.172<0,
\]
so by Result~\ref{thm:Q_controls_q0}\textup{(i)}, \(\log q_t\) decreases linearly with slope \(Q(e_1,e_2)\) on nonexit, that is, \(q_t\) decays exponentially; the trajectories that remain near resident fixation follow the dashed curve \(\log q_0+Q(e_1,e_2)\,t\). Five of the 20 trajectories leave the neighborhood early and settle at the interior equilibrium \(q=1/2\), as allowed by the local nature of Result~\ref{thm:Q_controls_q0}\textup{(i)}. Panels~\textup{(b,c)} use the cyclic game in Section~\ref{ex:rps} with \(b_0=0.9\) and \(\sigma=0.9\), resident \(x^\ast=\1/3\), and mutant \(e_1\), for which \(d=\|e_1-x^\ast\|^2=2/3\). The threshold is
\[
d_0^\ast(0.9)=\frac{\sqrt{12.24}-3}{2}\approx0.249.
\]
In panel~\textup{(b)}, \(d_0=0.6>d_0^\ast(0.9)\), so the resident lies above the threshold, with \(H(e_1,x^\ast)=1/9>0\). By Result~\ref{thm:SES_sufficient}\textup{(i)}, \(tq_t\to1/H(e_1,x^\ast)=9\) on nonexit, so \(q_t\) decays like \(9/t\), and the trajectories of \(tq_t\) approach this level. In panel~\textup{(c)}, \(d_0=d_0^\ast(0.9)\), so the resident lies exactly at the threshold, with \(H(e_1,x^\ast)=0\) and \(K(e_1,x^\ast)=V(e_1,x^\ast)/2\approx0.038>0\). The resident is still SES, but by Result~\ref{thm:zeroH_positiveK}\textup{(ii)}, \(\log q_t/\log t\to-1/2\) on nonexit, so \(q_t=t^{-1/2+o(1)}\) decays much more slowly than \(9/t\). Mutant frequency also fluctuates more, because the drift and the noise of its log-increments vanish at the same rate. The dashed line in panel~\textup{(c)} has log--log slope \(-1/2\).

\begin{figure*}[t]
\centering
\includegraphics[width=\textwidth]{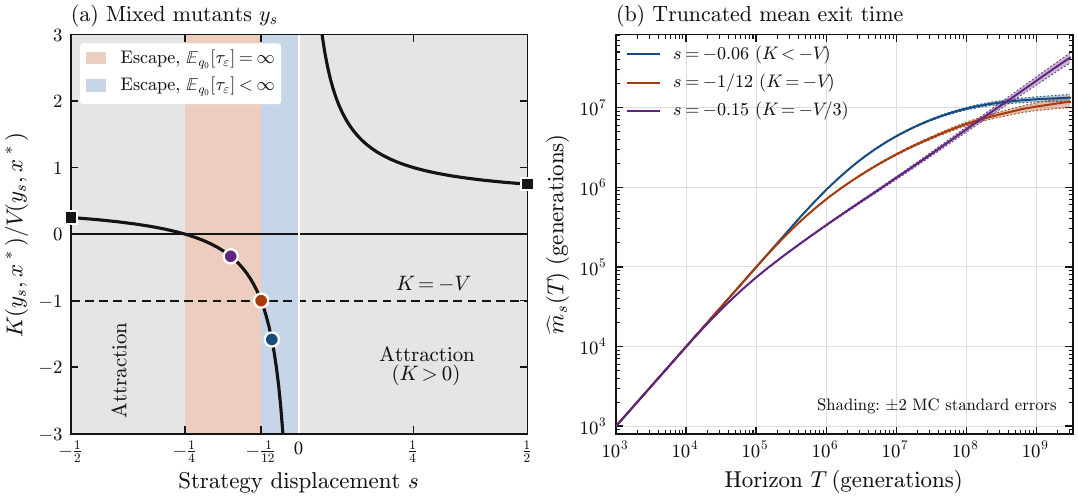}
\caption{Local escape and truncated mean exit times in the two-strategy game \eqref{eq:two_strategy_counterexample_matrices} of Section~\ref{ex:pure_mutants_not_ses}, with resident \(x^\ast=(1/2,1/2)^{\T}\) and mutants \(y_s=(1/2+s,1/2-s)^{\T}\). (a) The solid curve shows \(K(y_s,x^\ast)/V(y_s,x^\ast)\) for \(s\ne0\), and the dashed line marks \(K(y_s,x^\ast)=-V(y_s,x^\ast)\). Gray regions denote attraction. In the shaded escape regions, escape from sufficiently small neighborhoods of resident fixation occurs a.s., with infinite mean exit time for \(-1/4\le s\le-1/12\) (orange shading) and finite mean exit time for \(-1/12<s<0\) (blue shading). Squares mark the two pure mutants, both attracted; circles mark the three mutants simulated in (b). (b) Estimated truncated mean exit time \(\widehat m_s(T)\) against the horizon \(T\), both in generations, for \(s=-0.06\) (blue, \(K/V=-19/12\), finite mean), \(s=-1/12\) (orange, \(K/V=-1\), infinite mean), and \(s=-0.15\) (purple, \(K/V=-1/3\), infinite mean). For each mutant, \(N_s=10^4\) independent runs start at \(q_0=0.02\) and stop when \(q_t\) first reaches or exceeds \(\varepsilon=0.025\), or at \(3\times10^9\) generations if exit has not occurred; the same runs give the estimate at every horizon. Until runs begin to exit, \(\widehat m_s(T)=T\) and the three curves coincide. The shaded bands, outlined by thin dashed lines, show \(\widehat m_s(T)\pm2\,\mathrm{SE}\) at each horizon, where \(\mathrm{SE}\) is the Monte Carlo standard error over the \(10^4\) runs (approximately pointwise \(95\%\) confidence intervals for \(\E_{q_0}[\tau_{\varepsilon,s}\wedge T]\)).}
\label{fig:example_exit_times}
\end{figure*}

Figure~\ref{fig:example_exit_times}\textup{(b)} illustrates the exit-time regimes of Section~\ref{ex:pure_mutants_not_ses} by simulation. Each curve corresponds to one fixed mutant \(y_s\). Write \(\tau_{\varepsilon,s}\) for its exit time and \(\tau_{\varepsilon,s}^{(i)}\) for the exit time in run \(i\). At each horizon \(T\), we estimate the truncated mean exit time by
\[
\widehat m_s(T):=\frac{1}{N_s}\sum_{i=1}^{N_s}\min\{\tau_{\varepsilon,s}^{(i)},T\},
\]
so that a run that exits by \(T\) contributes its exit time and a run that has not yet exited contributes \(T\). The estimate approximates
\[
\E_{q_0}[\tau_{\varepsilon,s}\wedge T]=\int_0^T\mathbb P_{q_0}(\tau_{\varepsilon,s}>t)\,dt,
\]
which increases to \(\E_{q_0}[\tau_{\varepsilon,s}]\) as \(T\to\infty\) by the monotone convergence theorem.

For the blue mutant, \(s=-0.06\), we have \(K(y_s,x^\ast)<-V(y_s,x^\ast)\), so \(\E_{q_0}[\tau_{\varepsilon,s}]<\infty\). For the orange mutant, \(s=-1/12\), we have \(K(y_s,x^\ast)=-V(y_s,x^\ast)\), and for the purple mutant, \(s=-0.15\), we have \(-V(y_s,x^\ast)<K(y_s,x^\ast)<0\); in both cases \(\E_{q_0}[\tau_{\varepsilon,s}]=\infty\). The simulated curves agree with these conclusions. Between \(T=3\times10^8\) and \(3\times10^9\), the blue estimate gains \(1.7\times10^6\) generations, less than in the previous decade, and only \(2\) of \(10^4\) runs have not exited, so \(\widehat m_s(T)\approx1.33\times10^7\) is close to its limit. Over the same decade, the orange estimate gains about \(3.6\times10^6\) generations, as much as in earlier decades, and the purple estimate gains about four times more than in the previous decade, so neither curve levels off. Finite simulations cannot prove these conclusions, and each curve must be read on its own. Since \(V(y_s,x^\ast)=4s^4\), the intrinsic time scale ranges from about \(5\times10^4\) generations (purple) to \(2\times10^6\) (blue), which is why, once runs begin to exit (\(T\gtrsim3\times10^4\)), the blue estimate lies above the orange one, although only the blue mean is finite.

\section{Discussion}
In this paper, we studied invasion resistance in an infinite population with two fixed phenotypes and independent, identically distributed environments. Each phenotype uses a pure or mixed strategy, and in each generation \(t\) the fitness of one strategy against another is given by a random matrix \(A(t)\) with positive and bounded entries. At an SNE \(x^\ast\), every best reply \(y\) satisfies \(y^{\T}A(t)x^\ast=(x^\ast)^{\T}A(t)x^\ast\) almost surely, so a tied mutant and the resident differ only in their interactions with mutants. Two coefficients measure this difference: the first frequency-dependent coefficient \(H(y,x^\ast)\), which is the mean advantage of the resident over the mutant when both meet mutants, and the second-order coefficient \(K(y,x^\ast)\), which compares how strongly these two fitnesses fluctuate. Together with the strict-sign criteria of \cite{ZhengLiLessardTao2017,FengMatrix2022}, our analysis of the case \(H(y,x^\ast)=0\) gives a necessary and sufficient condition for SES (Result~\ref{cor:zeroH_SES}): \(x^\ast\) is SES if and only if every distinct tied mutant has either \(H(y,x^\ast)>0\), or \(H(y,x^\ast)=0\) and \(K(y,x^\ast)>0\). Along a feasible direction \(h=y-x^\ast\), this condition depends only on the signs of \(\E[C_h]\), \(\E[B_hC_h]\), and \(\E[C_h^2]\) (Result~9). It holds for any finite number of pure strategies, including residents on the boundary of the simplex.

The first results fix the setting for this condition. An SNE exists for any finite number of pure strategies (Result~2), and its best replies are exactly the strategies whose realized fitness against \(x^\ast\) equals the resident's own, almost surely. This tie property classifies SNEs by their best-reply sets (Result~3). An SNE is strong when it is its own unique best reply and fully weak when every strategy is a best reply; with two pure strategies every SNE is one of these two types, whereas with three or more, intermediate best-reply sets occur. A mutant that is not a best reply has a negative invasion exponent \(Q(y,x^\ast)\), defined as the expected logarithm of the mutant's fitness against the resident minus the expected logarithm of the resident's own fitness. For such mutants, Result~\ref{thm:Q_controls_q0} gives local stability, exponential extinction on nonexit, and an explicit bound on the probability of escape. Only tied mutants, for which \(Q(y,x^\ast)=0\), require a finer analysis.

The degenerate case has a simple interpretation. Write \(\alpha(t)\) for the fitness of the mutant against the mutant and \(\beta(t)\) for that of the resident against the mutant, both divided by the resident fitness \((x^\ast)^{\T}A(t)x^\ast\). Then \(H(y,x^\ast)=\E[\beta(t)-\alpha(t)]\), \(K(y,x^\ast)=\E[\alpha(t)^2-\beta(t)^2]/2\), and \(V(y,x^\ast)=\E[(\alpha(t)-\beta(t))^2]\), which measures the size of the fluctuations in their difference. When \(H(y,x^\ast)=0\), the two fitnesses have the same mean, and \(K(y,x^\ast)=(\operatorname{Var}\alpha(t)-\operatorname{Var}\beta(t))/2\). Near resident fixation, the mean logarithmic change of the odds is \(-K(y,x^\ast)z_t^2+O(z_t^3)\), and its variance is \(V(y,x^\ast)z_t^2+O(z_t^3)\). Here the odds \(z_t=q_t/(1-q_t)\) is the frequency of the mutant divided by the frequency of the resident. This shows that drift and noise are of the same order, and neither decides the outcome alone. In the reciprocal odds \(w_t=1/z_t\), the mean increment is \(\bigl(K(y,x^\ast)+V(y,x^\ast)/2\bigr)/w_t+O(w_t^{-2})\) and the mean squared increment tends to \(V(y,x^\ast)\). The mutant is attracted to extinction exactly when \(K(y,x^\ast)>0\), that is, when its fitness against mutants varies more than the resident's. If \(V(y,x^\ast)>0\) and \(K(y,x^\ast)\le0\), the mutant escapes locally almost surely. If \(V(y,x^\ast)=0\), then \(\alpha(t)=\beta(t)\) almost surely and the mutant frequency stays constant.

The examples show that the mean game can give the wrong conclusion. In a cyclic game, the first frequency-dependent coefficient vanishes only at a threshold of the noise level. The resident is still SES at this threshold, while below it fluctuations destroy the resistance of an ESS of the mean game. In a game whose mean is neutral, this coefficient vanishes for every mutant, and fluctuations alone create a unique SNE that is fully weak and SES. This settles a case left open in \cite{LiNoiseSNE2025} and shows that the strict condition of \cite{ZhengTaoLi2025} is not necessary. In the third example, a two-strategy game, both pure mutants are resisted, while mixed mutants arbitrarily close to the resident escape. Testing pure mutants is therefore not enough to establish stochastic evolutionary stability.

The extinction rates show how fast resistance acts. On nonexit, \(\log q_t/t\to Q(y,x^\ast)\) when \(Q(y,x^\ast)<0\); \(t\,q_t\to1/H(y,x^\ast)\) for a tied mutant with \(H(y,x^\ast)>0\); and \(\log q_t/\log t\to-1/2\) for a tied mutant with \(H(y,x^\ast)=0\) and \(K(y,x^\ast)>0\). The last limit gives the exponent of \(t\) but not the prefactor. In all stable cases, the probability of nonexit tends to one as \(q_0\to0\). A decline like \(t^{-1/2}\) is slow, and over a finite observation period it may be hard to tell apart from neutrality.

Exit times separate the unstable cases. For a tied mutant with \(H(y,x^\ast)=0\), \(V(y,x^\ast)>0\), and \(K(y,x^\ast)\le0\), exit occurs almost surely. Its expected time is finite if \(K(y,x^\ast)<-V(y,x^\ast)\) and infinite if \(-V(y,x^\ast)\le K(y,x^\ast)\le0\); the endpoint \(K(y,x^\ast)=-V(y,x^\ast)\) needs a separate argument. If \(\nu_1(w)\) and \(\nu_2(w)\) denote the mean and the mean square of the increment \(w_{t+1}-w_t\) given \(w_t=w\), the threshold is the sign of \(2w\nu_1(w)+\nu_2(w)=2\bigl(K(y,x^\ast)+V(y,x^\ast)\bigr)+O(w^{-1})\), which is Lamperti's criterion for passage-time moments of processes with vanishing drift \cite{Lamperti1963,DenisovKorshunovWachtel2025}. An unstable resident can therefore persist near fixation for a very long time, so finite observations may not reveal its instability.

Our results have two limitations. First, they concern one resident and one fixed mutant at a time. The size of the neighborhood and the rates depend on the mutant. Even at a strong SNE, \(Q(y,x^\ast)\to0\) as \(y\to x^\ast\), so no invasion barrier holds uniformly over all mutants. Second, the population is infinite and the environments are independent across generations. In a finite population, random sampling may dominate the slow regimes found here, and correlated environments may change the coefficients that decide stability.

\appendix

\section{Strict-sign invasion estimates}\label{app:strict_sign_invasion} 
Throughout the appendices, we use \(\mathbb P_{q_0}\) and \(\E_{q_0}\) for probabilities and expectations involving the population process with deterministic initial frequency \(q_0\). We retain \(\mathbb P\) and \(\E\) for probabilities and expectations involving only the environmental distribution.

Set \(R:=M/m>1\) and \(L:=\log R\). The odds corresponding to the exit threshold are
\[
z_\varepsilon:=\frac{\varepsilon}{1-\varepsilon}.
\]
By the fitness bounds, we have
\[
R^{-1}\le\frac{\pi_y(q,A)}{\pi_x(q,A)}\le R
\quad\text{and}\;\;
|\ell(q,A)|\le L,
\]
for $q\in[0,1]$. It follows that
\[
z_0R^{-t}\le z_t\le z_0R^t\quad\text{a.s.}
\]
These bounds also hold for \(z_{t\wedge\tau_\varepsilon}\). Then, every fixed positive or negative power of the stopped odds is integrable at each finite time.

We first establish two estimates for stopped processes that will be used in the proofs below.

\begin{lemma}\label{lem:stopped_drift_estimates}
Fix \(0<q_0<\varepsilon<1\).
\begin{enumerate}
\item[(i)] Suppose that, for some \(\theta>0\), \(c>0\), and \(\gamma\ge0\),
\[
\E_{q_0}[z_{t+1}^{\theta}-z_t^{\theta}\mid\mathcal F_t]
\le-cz_t^{\theta+\gamma}
\quad\text{on }\{t<\tau_\varepsilon\}.
\]
Then
\[
\begin{aligned}
&\mathbb P_{q_0}(\tau_\varepsilon<\infty)\le\left(\frac{z_0}{z_\varepsilon}\right)^\theta,\\
&\E_{q_0}\!\left[\sum_{t=0}^{\infty}z_t^{\theta+\gamma}\mathbf1_{\{t<\tau_\varepsilon\}}\right]\le\frac{z_0^\theta}{c}.
\end{aligned}
\]
In particular, \(q_t\to0\) a.s. on \(\{\tau_\varepsilon=\infty\}\).

\item[(ii)] Let \((X_t)\) be a nonnegative adapted process that is integrable at every finite time. If, for some \(c>0\),
\[
\E_{q_0}[X_{t+1}-X_t\mid\mathcal F_t]\le-c
\quad\text{on }\{t<\tau_\varepsilon\},
\]
then
\[
\E_{q_0}[\tau_\varepsilon]\le\frac{\E_{q_0}[X_0]}{c}.
\]
\end{enumerate}
\end{lemma}

\begin{proof}
First, note 
\begin{equation}\label{eq:stopped_increment_identity}
U_{(t+1)\wedge\tau_\varepsilon}-U_{t\wedge\tau_\varepsilon}
=\mathbf1_{\{t<\tau_\varepsilon\}}(U_{t+1}-U_t),
\end{equation}
for any process \((U_t)\), where \(\mathbf1_{\{t<\tau_\varepsilon\}}\) is \(\mathcal F_t\)-measurable.

\textbf{For part~\textup{(i)}}, applying \eqref{eq:stopped_increment_identity} with \(U_t=z_t^\theta\) and using the assumed conditional drift bound, we have
\[
\begin{aligned}
\E_{q_0}[z_{(t+1)\wedge\tau_\varepsilon}^{\theta}\mid\mathcal F_t]&=z_{t\wedge\tau_\varepsilon}^{\theta}+\E_{q_0}\left[\mathbf1_{\{t<\tau_\varepsilon\}}(z_{t+1}^\theta-z_t^\theta)\mid\mathcal F_t\right]\\
&\le z_{t\wedge\tau_\varepsilon}^{\theta} -cz_t^{\theta+\gamma}\mathbf1_{\{t<\tau_\varepsilon\}}.
\end{aligned}
\]
Taking expectations and summing over \(t=0,\ldots,N-1\), we obtain
\begin{equation}\label{eq_appendix_A_eq1}
\E_{q_0}[z_{N\wedge\tau_\varepsilon}^{\theta}]+c\E_{q_0}\!\left[\sum_{t=0}^{N-1}z_t^{\theta+\gamma}\mathbf1_{\{t<\tau_\varepsilon\}}\right]\le z_0^\theta.
\end{equation}
On \(\{\tau_\varepsilon\le N\}\), we have \(z_{N\wedge\tau_\varepsilon}^{\theta}\ge z_\varepsilon^\theta\). Consequently,
\[
z_\varepsilon^\theta\mathbb P_{q_0}(\tau_\varepsilon\le N)
\le\E_{q_0}[z_{N\wedge\tau_\varepsilon}^{\theta}]
\le z_0^\theta.
\]
The last inequality is a consequence of \eqref{eq_appendix_A_eq1}. Letting \(N\to\infty\) gives the stated bound on the exit probability.

Since the first term in \eqref{eq_appendix_A_eq1} is nonnegative, the monotone convergence theorem also gives
\[
\E_{q_0}\!\left[
\sum_{t=0}^{\infty}
z_t^{\theta+\gamma}\mathbf1_{\{t<\tau_\varepsilon\}}
\right]
\le\frac{z_0^\theta}{c}.
\]
Thus the sum $\sum_{t=0}^{\infty}z_t^{\theta+\gamma}\mathbf1_{\{t<\tau_\varepsilon\}}$ is finite a.s. In particular, on \(\{\tau_\varepsilon=\infty\}\), its summands $z_t^{\theta+\gamma}\mathbf1_{\{t<\tau_\varepsilon\}}=z_t^{\theta+\gamma}$ tend to zero. Since \(\theta+\gamma>0\), we obtain \(z_t\to0\), and hence \(q_t\to0\), a.s.

\textbf{For part~\textup{(ii)}}, and similarly, applying \eqref{eq:stopped_increment_identity} with \(U_t=X_t\) and using the assumed drift inequality, we obtain 
\[
\E_{q_0}[X_{(t+1)\wedge\tau_\varepsilon}\mid\mathcal F_t]\le X_{t\wedge\tau_\varepsilon}-c\mathbf1_{\{t<\tau_\varepsilon\}}.
\]
Taking expectations and summing over \(t=0,\ldots,N-1\) gives
\[
\E_{q_0}[X_{N\wedge\tau_\varepsilon}]
+c\E_{q_0}[N\wedge\tau_\varepsilon]
\le\E_{q_0}[X_0],
\]
where we used
\[
\sum_{t=0}^{N-1}\mathbf1_{\{t<\tau_\varepsilon\}}
=N\wedge\tau_\varepsilon.
\]
Since \(X_{N\wedge\tau_\varepsilon}\ge0\), we have
\[
c\E_{q_0}[N\wedge\tau_\varepsilon]\le\E_{q_0}[X_0].
\]
The monotone convergence theorem then yields
\[
\E_{q_0}[\tau_\varepsilon]\le\frac{\E_{q_0}[X_0]}{c},
\]
which completes the proof.
\end{proof}

\begin{remark}\label{remark_appendixA_1}
Suppose that the drift condition in Lemma \ref{lem:stopped_drift_estimates}~\textup{(i)} holds for every sufficiently small \(\varepsilon\), with constants independent of \(q_0\in(0,\varepsilon)\). The lemma then implies local asymptotic stability in probability. Indeed, for any \(\eta\in(0,1)\), choosing
\[
0<q_0<
\frac{\eta^{1/\theta}z_\varepsilon}
{1+\eta^{1/\theta}z_\varepsilon}
\]
gives \(\mathbb P_{q_0}(\tau_\varepsilon<\infty)<\eta\). Since \(q_t\to0\) a.s. on \(\{\tau_\varepsilon=\infty\}\), the probability that the mutant frequency remains below \(\varepsilon\) and converges to zero is at least \(1-\eta\).
\end{remark}

\begin{remark}\label{remark_appendixA_2}
We also use the strong law for martingales with bounded increments. Let \((M_t)\) be a martingale such that \(M_0=0\) and \(|M_{t+1}-M_t|\le C\). Then
\[
\sum_{t=0}^{\infty}
\frac{\E_{q_0}[(M_{t+1}-M_t)^2]}{(t+1)^2}
\le C^2\sum_{t=0}^{\infty}\frac1{(t+1)^2}
<\infty.
\]
Then, the martingale
\[
\widetilde M_N
:=\sum_{t=0}^{N-1}\frac{M_{t+1}-M_t}{t+1}
\]
is bounded in \(L^2\). By the martingale convergence theorem in $L^2$, \(\widetilde M_N\) converges a.s. Applying Kronecker's lemma, we obtain
\begin{equation}\label{eq:appendix_martingale_strong_law}
\frac{1}{t}\sum_{t'=0}^{t-1}\left(M_{t'+1}-M_{t'}\right)=\frac{M_t}{t}\longrightarrow0
\quad\text{a.s.}
\end{equation}
\end{remark}

\subsection{A nonzero boundary invasion exponent}
\label{app:local_invasion}

Define \(\mu(q):=\E[\ell(q,A(0))]\) for \(q\in[0,1]\). Since \(\ell(q,A)\) is continuous in \(q\) and uniformly bounded, the dominated convergence theorem implies that \(\mu\) is continuous. Moreover, \(\mu(0)=Q(y,x)\).

\begin{lemma}[Uniform exponential estimate]
\label{lem:strict_sign_exponential_estimate}
Suppose that \(Q(y,x)\ne0\), and let \(\sigma=-sgn(Q(y,x))\in\{-1,1\}\). Then, there exist \(\theta>0\), \(r\in(0,1)\), and \(\varepsilon_0\in(0,1)\) such that
\begin{equation}\label{eq:strict_sign_exponential_estimate}
\E\!\left[e^{\sigma\theta\ell(q,A(0))}\right]\le r
\quad\text{for every }q\in[0,\varepsilon_0].
\end{equation}
\end{lemma}

\begin{proof}
Define
\[
G_\sigma(\theta,q)
:=\E\!\left[e^{\sigma\theta\ell(q,A(0))}\right].
\]
Since \(\ell\) is uniformly bounded, differentiation under the expectation gives
\[
G_\sigma(0,0)=1,
\quad
\left.\frac{\partial G_\sigma}{\partial\theta}\right|_{(0,0)}
=\sigma Q(y,x)=-|Q(y,x)|<0.
\]
It follows that \(G_\sigma(\theta,0)<1\) for every sufficiently small \(\theta>0\). Fix such a \(\theta\), and choose \(r\) satisfying
\[
G_\sigma(\theta,0)<r<1.
\]
By the dominated convergence theorem, \(G_\sigma(\theta,q)\) is continuous in \(q\). Therefore, there exists \(\varepsilon_0\in(0,1)\) such that \(G_\sigma(\theta,q)\le r\) for every \(q\in[0,\varepsilon_0]\), as required.
\end{proof}
\begin{proof}[\textbf{Proof of Result~\ref{thm:Q_controls_q0}\textup{(i)}}]
Suppose that \(Q(y,x)<0\). By Lemma~\ref{lem:strict_sign_exponential_estimate} with \(\sigma=1\), there exist \(\theta>0\), \(r\in(0,1)\), and \(\varepsilon_0\in(0,1)\) satisfying \eqref{eq:strict_sign_exponential_estimate}. Fix \(0<q_0<\varepsilon\le\varepsilon_0\). Since \(q_t\) and \(z_t\) are \(\mathcal F_t\)-measurable and \(A(t)\) is independent of \(\mathcal F_t\), we have, on \(\{t<\tau_\varepsilon\}\),
\[
\E_{q_0}[z_{t+1}^\theta\mid\mathcal F_t]
=z_t^\theta\E_{q_0}\!\left[e^{\theta\ell(q_t,A(t))}\mid\mathcal F_t\right]
\le r z_t^\theta .
\]
Applying Lemma~\ref{lem:stopped_drift_estimates}\textup{(i)} with \(c=1-r\) and \(\gamma=0\) gives
\begin{equation}\label{eq:exit_probability_negative_Q}
\mathbb P_{q_0}(\tau_\varepsilon<\infty)
\le\left(\frac{z_0}{z_\varepsilon}\right)^\theta
\end{equation}
and \(q_t\to0\) a.s. on \(\{\tau_\varepsilon=\infty\}\). Remark~\ref{remark_appendixA_1} implies that \(\hat q=0\) is locally asymptotically stable in probability against \(y\).

To determine the extinction rate, define
\[
M_t:=\sum_{s=0}^{t-1}\bigl(\ell(q_s,A(s))-\mu(q_s)\bigr).
\]
Since \(q_s\) is \(\mathcal F_s\)-measurable and \(A(s)\) is independent of \(\mathcal F_s\), conditioning on the past fixes \(q_s\) and averages over the current environment. Hence
\[
\E_{q_0}[\ell(q_s,A(s))\mid\mathcal F_s]=\mu(q_s),
\]
so each increment of \(M_t\) has conditional mean zero. Moreover, \(|\ell(q,A)|\le L\) implies \(|\mu(q)|\le L\), giving \(|M_{s+1}-M_s|\le2L\). Thus \((M_t)\) is a martingale with uniformly bounded increments, and \eqref{eq:appendix_martingale_strong_law} yields
\begin{equation}\label{eq_appendix_A_eq2}
\frac{M_t}{t}\longrightarrow0
\quad\text{a.s.}
\end{equation}
On the other hand, summing \eqref{eq:log_odds_recursion} and dividing by \(t\), we obtain
\begin{equation}\label{eq_appendix_A_eq3}
\frac{\log z_t}{t}
=\frac{\log z_0}{t}
+\frac{M_t}{t}
+\frac1t\sum_{s=0}^{t-1}\mu(q_s).
\end{equation}
On \(\{\tau_\varepsilon=\infty\}\), we have \(q_s\to0\) a.s. Continuity of \(\mu\) at zero therefore gives \(\mu(q_s)\to\mu(0)=Q(y,x)\), and Cesàro's theorem implies
\begin{equation}\label{eq_appendix_A_eq4}
\frac1t\sum_{s=0}^{t-1}\mu(q_s)\longrightarrow Q(y,x)
\quad\text{a.s. on }\{\tau_\varepsilon=\infty\}.
\end{equation}
Combining \eqref{eq_appendix_A_eq2}--\eqref{eq_appendix_A_eq4} yields
\[
\frac{\log z_t}{t}\longrightarrow Q(y,x)
\quad\text{a.s. on }\{\tau_\varepsilon=\infty\}.
\]
Finally, since \(\log q_t=\log z_t-\log(1+z_t)\), and \(z_t\to0\) a.s. on the same event, we obtain
\[
\lim_{t\to\infty}\frac{\log q_t}{t}=Q(y,x)
\quad\text{a.s. on }\{\tau_\varepsilon=\infty\}.
\]
\end{proof}

\begin{proof}[\textbf{Proof of Result~\ref{thm:Q_controls_q0}\textup{(ii)}}]
Suppose that \(Q(y,x)>0\). By Lemma~\ref{lem:strict_sign_exponential_estimate} with \(\sigma=-1\), there exist \(\theta>0\), \(r\in(0,1)\), and \(\varepsilon_0\in(0,1)\) such that, for \(0<q_0<\varepsilon\le\varepsilon_0\),
\[
\E_{q_0}[z_{t+1}^{-\theta}\mid\mathcal F_t]
\le r z_t^{-\theta}
\quad\text{on }\{t<\tau_\varepsilon\}.
\]
Define
\begin{equation}\label{eq:strict_sign_stopped_process}
Y_t:=r^{-(t\wedge\tau_\varepsilon)}
z_{t\wedge\tau_\varepsilon}^{-\theta}.
\end{equation}
On \(\{t<\tau_\varepsilon\}\), we have \((t+1)\wedge\tau_\varepsilon=t+1\), and hence
\[
\E_{q_0}[Y_{t+1}\mid\mathcal F_t]
=r^{-(t+1)}
\E_{q_0}[z_{t+1}^{-\theta}\mid\mathcal F_t]
\le r^{-t}z_t^{-\theta}
=Y_t.
\]
On \(\{t\ge\tau_\varepsilon\}\), \(Y_{t+1}=Y_t\). Therefore, \((Y_t)\) is a nonnegative supermartingale, and
\begin{equation}\label{eq:strict_sign_stopped_expectation}
\E_{q_0}[Y_t]\le z_0^{-\theta}
\quad\text{for all}\;t\ge0.
\end{equation}

Before exit, we have \(z_t<z_\varepsilon\). If exit occurs, then \(\tau_\varepsilon\ge1\), and the bound \(\dfrac{\pi_y(q,A)}{\pi_x(q,A)}\le R\) gives
\[
z_{\tau_\varepsilon}
=z_{\tau_\varepsilon-1}
\frac{\pi_y(q_{\tau_\varepsilon-1},A(\tau_\varepsilon-1))}{\pi_x(q_{\tau_\varepsilon-1},A(\tau_\varepsilon-1))}
\le Rz_\varepsilon.
\]
It follows that \(z_{N\wedge\tau_\varepsilon}\le Rz_\varepsilon\) for every \(N\ge0\). Combining this bound with \eqref{eq:strict_sign_stopped_expectation}, we obtain
\[
(Rz_\varepsilon)^{-\theta}
\E_{q_0}\!\left[r^{-(N\wedge\tau_\varepsilon)}\right]
\le\E_{q_0}[Y_N]\le z_0^{-\theta}.
\]
Set \(\lambda:=-\log r>0\). Since \(e^{\lambda(N\wedge\tau_\varepsilon)}\) increases to \(e^{\lambda\tau_\varepsilon}\) as $N$ increases, the monotone convergence theorem gives
\begin{equation}\label{eq:positive_Q_exponential_moment_bound}
\E_{q_0}\!\left[e^{\lambda\tau_\varepsilon}\right]
\le\left(\frac{Rz_\varepsilon}{z_0}\right)^\theta
<\infty.
\end{equation}
In particular, \(\tau_\varepsilon<\infty\) a.s. This holds for every sufficiently small \(\varepsilon\) and every \(q_0\in(0,\varepsilon)\). Therefore, the resident fixation state is not locally stable in probability, which completes the proof.
\end{proof}

\subsection{A nonzero first frequency-dependent coefficient}
\label{app:tied_first_order}

Let \(x^\ast\) be an SNE and fix \(y\in\BR(x^\ast)\setminus\{x^\ast\}\). Define \(a:=y^{\T}A(t)y\), \(b:=(x^\ast)^{\T}A(t)y\), and \(c:=(x^\ast)^{\T}A(t)x^\ast\). Tie rigidity gives \(y^{\T}A(t)x^\ast=c\) a.s.
The exact recursion gives
\[
z_{t+1}=z_t\frac{\pi_y(q_t,A(t))}{\pi_x(q_t,A(t))},
\qquad
w_{t+1}-w_t=D(q_t,A(t)),
\]
where
\begin{equation}\label{eq:reciprocal_odds_increment}
\begin{aligned}
\frac{\pi_y(q,A)}{\pi_x(q,A)}
&=1+\frac{q(a-b)}{c+q(b-c)},\\
D(q,A)
&=\frac{(1-q)(b-a)}{c+q(a-c)}.
\end{aligned}
\end{equation}
Both denominators lie in \([m,M]\) for \(q\in[0,1]\). Consequently, we have
\begin{equation}\label{eq:tied_bounded_reciprocal_increment}
|D(q,A)|\le B:=\frac{M-m}{m},
\quad\text{for}\;\; 0\le q\le1.
\end{equation}

Define
\[
h(q):=\E\!\left[\frac{b-a}{c+q(b-c)}\right],
\qquad
d(q):=\E[D(q,A)],
\]
where the expectations are taken over the environmental matrix \(A\). The integrands are continuous in \(q\) and uniformly bounded, so the dominated convergence theorem implies that \(h\) and \(d\) are continuous on \([0,1]\). Moreover,
\[
h(0)=d(0)=\E\!\left[\frac{b-a}{c}\right]=H(y,x^\ast).
\]
Since \(q_t\), \(z_t\), and \(w_t\) are \(\mathcal F_t\)-measurable and \(A(t)\) is independent of \(\mathcal F_t\), we obtain
\begin{equation}\label{eq:tied_conditional_drifts}
\begin{aligned}
\E_{q_0}[z_{t+1}\mid\mathcal F_t]
&=z_t\bigl(1-q_t h(q_t)\bigr),\\
\E_{q_0}[w_{t+1}-w_t\mid\mathcal F_t]
&=d(q_t).
\end{aligned}
\end{equation}

\begin{proof}[\textbf{Proof of Result~\ref{thm:SES_sufficient}\textup{(i)}}]
Suppose that \(H(y,x^\ast)>0\). By continuity of \(h\) at $0$, there exists \(\varepsilon_0\in(0,1)\) such that \(h(q)\ge H(y,x^\ast)/2\) for every \(q\in[0,\varepsilon_0]\). Fix \(0<q_0<\varepsilon\le\varepsilon_0\). On \(\{t<\tau_\varepsilon\}\), we have \(z_t<z_\varepsilon\) and \(q_t=z_t/(1+z_t)\). Hence \eqref{eq:tied_conditional_drifts} gives
\[
\begin{aligned}
\E_{q_0}[z_{t+1}-z_t\mid\mathcal F_t]
&=-z_tq_t h(q_t)\\
&\le-\frac{H(y,x^\ast)}{2}\frac{z_t^2}{1+z_t}\\
&\le-\frac{H(y,x^\ast)}{2(1+z_\varepsilon)}z_t^2.
\end{aligned}
\]
Applying Lemma~\ref{lem:stopped_drift_estimates}\textup{(i)} with \(\theta=\gamma=1\) and drift constant \(H(y,x^\ast)/[2(1+z_\varepsilon)]\), we obtain
\begin{equation}\label{eq:tied_exit_probability}
\mathbb P_{q_0}(\tau_\varepsilon<\infty)
\le\frac{z_0}{z_\varepsilon}
\end{equation}
and \(q_t\to0\) a.s. on \(\{\tau_\varepsilon=\infty\}\). Remark \ref{remark_appendixA_1} shows the local asymptotic stability in probability of $\hat q=0$ against $y$.

To determine the extinction rate, define
\[
N_t:=\sum_{s=0}^{t-1}\mathbf1_{\{s<\tau_\varepsilon\}}
\bigl(w_{s+1}-w_s-d(q_s)\bigr).
\]
The indicator \(\mathbf1_{\{s<\tau_\varepsilon\}}\) is \(\mathcal F_s\)-measurable, so \eqref{eq:tied_conditional_drifts} implies that each increment of \(N_t\) has conditional mean zero. Moreover, \eqref{eq:tied_bounded_reciprocal_increment} gives \(|d(q)|\le B\), and hence \(|N_{s+1}-N_s|\le2B\). Thus \((N_t)\) is a martingale with uniformly bounded increments, and \eqref{eq:appendix_martingale_strong_law} yields 
\begin{equation}\label{eq_appendix_A_eq5}
\frac{N_t}{t}\longrightarrow0
\quad\text{a.s.}
\end{equation}

Using \eqref{eq:stopped_increment_identity} for $w_s$, and summing for $0,1,\ldots,t-1$, we obtain
\begin{equation}\label{eq:tied_reciprocal_decomposition}
\begin{aligned}
w_{t\wedge\tau_\varepsilon}&=w_0+\sum_{s=0}^{t-1}\mathbf1_{\{s<\tau_\varepsilon\}}(w_{s+1}-w_s)  \\
&=w_0+N_t+\sum_{s=0}^{t-1}\mathbf1_{\{s<\tau_\varepsilon\}}d(q_s).
\end{aligned}
\end{equation}
On \(\{\tau_\varepsilon=\infty\}\), we have \(q_s\to0\) a.s., so continuity of \(d\) at $0$ gives 
\[
d(q_s)\to d(0)=H(y,x^\ast).
\]
Dividing \eqref{eq:tied_reciprocal_decomposition} by \(t\) and applying Cesàro's theorem, we conclude that
\[
\frac{w_t}{t}\longrightarrow H(y,x^\ast)
\quad\text{a.s. on }\{\tau_\varepsilon=\infty\}.
\]
As \(q_t=1/(1+w_t)\), we deduce that
\[
tq_t\longrightarrow\frac{1}{H(y,x^\ast)}
\quad\text{a.s. on }\{\tau_\varepsilon=\infty\}.
\]
\end{proof}

\begin{proof}[\textbf{Proof of Result~\ref{thm:SES_sufficient}\textup{(ii)}}]
Suppose that \(H(y,x^\ast)<0\). By continuity of \(d\) at $0$, there exists \(\varepsilon_0\in(0,1)\) such that
\[
d(q)\le\frac{d(0)}{2}=-\frac{|H(y,x^\ast)|}{2}
\quad\text{for every }q\in[0,\varepsilon_0].
\]
Fix \(0<q_0<\varepsilon\le\varepsilon_0\). Equation~\eqref{eq:tied_conditional_drifts} gives
\[
\E_{q_0}[w_{t+1}-w_t\mid\mathcal F_t]
\le-\frac{|H(y,x^\ast)|}{2}
\quad\text{on }\{t<\tau_\varepsilon\}.
\]
By \eqref{eq:tied_bounded_reciprocal_increment}, we have 
\[0<w_t=w_0+\sum_{s=1}^{t}(w_s-w_{s-1})\le w_0+tB,\] so the reciprocal odds are integrable at every finite time. Applying Lemma~\ref{lem:stopped_drift_estimates}\textup{(ii)} with \(X_t=w_t\) and drift constant \(|H(y,x^\ast)|/2\), we obtain
\begin{equation}\label{eq:tied_mean_exit_bound}
\E_{q_0}[\tau_\varepsilon]
\le\frac{2w_0}{|H(y,x^\ast)|}<\infty.
\end{equation}
Consequently, exit occurs a.s., and the resident fixation state is not locally stable in probability against \(y\).
\end{proof}
\section{Proof of Results~\ref{prop:SNE_existence} and \ref{res:strong_fully_weak}}
\label{app:SNE_structure}

\subsection{Existence of a stochastic Nash equilibrium in Result~\ref{prop:SNE_existence}}

Assumption \textup{($\mathcal A2$)} gives
\[
\left|\log\bigl(y^{\T}A(t)x\bigr)\right|
\le \max\{|\log m|,|\log M|\}
\]
uniformly over the simplex. Since the integrand is continuous in \((x,y)\) for almost every realization of \(A(t)\), the dominated convergence theorem shows that \((x,y)\mapsto F(x,y)\) is continuous.

Fix \(x\in\Delta_{n-1}\). Continuity of \(F(x,\cdot)\) and compactness of \(\Delta_{n-1}\) ensure that the maximum is attained, so \(\BR(x)\) is nonempty. The set of maximizers is closed and therefore compact. Moreover, \(F(x,\cdot)\) is concave because the logarithm is concave and \(y^{\T}A(t)x\) is affine in \(y\). Every convex combination of two maximizers consequently has a value at least as large as their common maximum, and hence is also a maximizer. Thus \(\BR(x)\) is convex.

To prove that at least one SNE exists, we apply Kakutani's fixed-point theorem \cite{Kakutani1941} to the correspondence
\[
\BR:\Delta_{n-1}\longrightarrow\mathcal P(\Delta_{n-1}),
\]
where \(\mathcal P(\Delta_{n-1})\) denotes the collection of all subsets of \(\Delta_{n-1}\). The theorem requires:
\begin{enumerate}
\item the domain to be a nonempty compact convex subset of a finite-dimensional Euclidean space;
\item each value to be a nonempty compact convex subset of the domain;
\item the correspondence to be upper hemicontinuous.
\end{enumerate}
The simplex satisfies the first condition, and the properties proved above give the second. Since the range is contained in the compact simplex and the values are closed, it remains to verify that the graph of \(\BR\) is closed. Suppose that \(x_k\to x\) and \(y_k\to y\), with \(y_k\in\BR(x_k)\) for every \(k\ge1\). For any fixed \(z\in\Delta_{n-1}\),
\[
F(x_k,y_k)\ge F(x_k,z)\quad\text{for every }k\ge1.
\]
Continuity of \(F\) allows passage to the limit, yielding
\[
F(x,y)\ge F(x,z).
\]
Since \(z\) was arbitrary, \(y\) maximizes \(F(x,\cdot)\). Thus \(y\in\BR(x)\), establishing the third condition.

By Kakutani's theorem, there exists at least one \(x^\ast\in\Delta_{n-1}\) with \(x^\ast\in\BR(x^\ast)\), which is an SNE.

\subsection{Tie rigidity in Result~\ref{prop:SNE_existence}}

Suppose that \(y\in\BR(x^\ast)\). Since \(x^\ast\) is an SNE, both \(x^\ast\) and \(y\) belong to \(\BR(x^\ast)\). Convexity of the best-reply set shows that their midpoint is also a maximizer. Therefore,
\begin{equation}\label{eq_equality_three_expected_logarithmic}
F\!\left(x^\ast,\frac{x^\ast+y}{2}\right)
=F(x^\ast,x^\ast)=F(x^\ast,y).
\end{equation}
Set \(u=(x^\ast)^{\T}A(t)x^\ast\) and \(v=y^{\T}A(t)x^\ast\). The midpoint has realized fitness \((u+v)/2\) against the resident. Since \(u,v>0\), strict concavity of the logarithm gives
\[
G:=\log\frac{u+v}{2}-\frac12\log u-\frac12\log v\ge0,
\]
with equality if and only if \(u=v\). Assumption \textup{($\mathcal A2$)} ensures that \(G\) is integrable, and \eqref{eq_equality_three_expected_logarithmic} gives
\[
\E[G]
=F\!\left(x^\ast,\frac{x^\ast+y}{2}\right)
-\frac12F(x^\ast,x^\ast)-\frac12F(x^\ast,y)
=0.
\]
Since \(G\) is nonnegative and has zero expectation, then \(G=0\) a.s. It follows that \(u=v\) a.s., proving \eqref{eq:tie_rigidity}.

Conversely, suppose that \eqref{eq:tie_rigidity} holds. Taking logarithms and expectations yields
\[
F(x^\ast,y)=F(x^\ast,x^\ast).
\]
Since \(x^\ast\) is an SNE, the right-hand side is the maximum of \(F(x^\ast,\cdot)\) over \(\Delta_{n-1}\). Thus \(y\in\BR(x^\ast)\).

\subsection{Proof of Result~\ref{res:strong_fully_weak}}
Let $T:=\{h\in\mathbb R^n:\1^{\T}h=0\}$. Every displacement between two strategies belongs to \(T\).

\textbf{For part~\textup{(i)}}, tie rigidity shows that \(x^\ast\) has no distinct tied best reply if and only if
\begin{equation}\label{eq:strong_SNE_characterization}
\mathbb P\!\left((y-x^\ast)^{\T}A(t)x^\ast\ne0\right)>0
\end{equation}
for every \(y\in\Delta_{n-1}\setminus\{x^\ast\}\). This proves the characterization of a strong SNE.

\textbf{For part~\textup{(ii)}}, write \(c(t):=(x^\ast)^{\T}A(t)x^\ast\). If \(x^\ast\) is fully weak, each pure strategy \(e_i\) is a best reply. Tie rigidity gives
\[
e_i^{\T}A(t)x^\ast=c(t)\quad\text{a.s.},
\qquad i=1,\ldots,n.
\]

Conversely, on the event where this vector identity holds, we obtain
\[
y^{\T}A(t)x^\ast=c(t)\1^{\T}y=c(t)
\qquad\text{for every }y\in\Delta_{n-1}.
\]
Tie rigidity therefore shows that every strategy is a best reply.

Finally, suppose that \(n=2\) and \(x^\ast\) is not strong. There is a distinct tied best reply \(y\), and \(h:=y-x^\ast\ne0\) satisfies
\[
h^{\T}A(t)x^\ast=0\quad\text{a.s.}
\]
Since \(T\) is one-dimensional, for every \(z\in\Delta_1\) there exists \(s\in\mathbb R\) such that \(z-x^\ast=sh\). On the event where \(h^{\T}A(t)x^\ast=0\), we have
\[
(z-x^\ast)^{\T}A(t)x^\ast
=s\,h^{\T}A(t)x^\ast=0
\qquad\text{for every }z\in\Delta_1.
\]
Tie rigidity implies that every strategy is a best reply. Thus \(x^\ast\) is fully weak.

\begin{remark}[Interior residents]
\label{rem:interior_strong_SNE}
Let \(x^\ast\) be an interior SNE. It is strong if and only if
\begin{equation}\label{eq:interior_strong_SNE}
\mathbb P\!\left(h^{\T}A(t)x^\ast\ne0\right)>0
\qquad\text{for every }h\in T\setminus\{0\}.
\end{equation}
Indeed, every nonzero \(h\in T\) gives a feasible mutant \(y=x^\ast+sh\) for sufficiently small \(s>0\), with
\[
(y-x^\ast)^{\T}A(t)x^\ast=s\,h^{\T}A(t)x^\ast.
\]
Conversely, every distinct mutant has a nonzero displacement in \(T\). The equivalence follows from \eqref{eq:strong_SNE_characterization}.
\end{remark}

\section{Stability with vanishing first-order drift}
\label{app:zeroH_stability}

Throughout this appendix, Assumptions \textup{($\mathcal A1$)}--\textup{($\mathcal A3$)} hold. Let \(x^\ast\) be an SNE and fix a distinct tied mutant \(y\in\BR(x^\ast)\setminus\{x^\ast\}\) satisfying \(H(y,x^\ast)=0\). Since \eqref{eq:zeroH_exact_neutrality} establishes exact neutrality when \(V(y,x^\ast)=0\), we assume \(V(y,x^\ast)>0\) below.

\subsection{Increment estimates}

Recall the reciprocal-odds increment \(\phi(w,A)\) defined in \eqref{eq:zeroH_reciprocal_increment}. By \eqref{eq:tied_bounded_reciprocal_increment}, we have \(|\phi(w,A)|\le B\). 

\begin{lemma}
\label{lem:zeroH_reciprocal}
The increment moments satisfy \eqref{eq:zeroH_increment_moments}. Moreover, for every fixed \(p\in\mathbb R\), we have
\begin{equation}\label{eq:zeroH_reciprocal_power_drift}
\begin{aligned}
&\E\!\left[(w+\phi(w,A))^p-w^p\right]\\
&=p\left(K(y,x^\ast)+\frac{pV(y,x^\ast)}{2}\right)w^{p-2}
+O(w^{p-3})
\end{aligned}
\end{equation}
as \(w\to\infty\).
\end{lemma}

\begin{proof}
Since \(\alpha\) and \(\beta\) are uniformly bounded, expanding \(\phi(w,A)\) gives
\begin{equation}\label{eqq3}
\phi(w,A)=\frac{w(\beta-\alpha)}{w+\alpha}=(\beta-\alpha)-\frac{\alpha(\beta-\alpha)}{w}+O(w^{-2}),
\end{equation}
where the remainder is uniform over environmental realizations. Taking expectations in the expansion and using the fact that $\E[\beta-\alpha]=H(y,x^\ast)=0$ and $\E[\alpha(\alpha-\beta)]=K(y,x^\ast)+V(y,x^\ast)/2$, we  obtain
\begin{equation}\label{eqq4}
\begin{aligned}
&\E\left[\phi(w,A)\right]\\
&=H(y,x^\ast)+\frac{1}{w}\left(K(y,x^\ast)+\frac{V(y,x^\ast)}{2}\right)+O(w^{-2})\\
&=\frac{K(y,x^\ast)+V(y,x^\ast)/2}{w}+O(w^{-2}).
\end{aligned}
\end{equation}
Similarly, squaring the expansion in \eqref{eqq3} yields
\[
\phi(w,A)^2=(\beta-\alpha)^2+O(w^{-1}).
\]
Hence, we obtain
\begin{equation}\label{eqq5}
\E\left[\phi(w,A)^2\right]=V(y,x^\ast)+O(w^{-1}),
\end{equation}
which proves \eqref{eq:zeroH_increment_moments}.

To obtain the power drift, let \(w>2B\). Since \(|\phi(w,A)/w|<1/2\), Taylor's theorem gives
\[
\begin{aligned}
&(w+\phi(w,A))^p-w^p\\
&=pw^{p-1}\phi(w,A)+\frac{p(p-1)}2w^{p-2}\phi(w,A)^2+R_p(w,A),
\end{aligned}
\]
where
\[
|R_p(w,A)|
\le \gamma_p w^{p-3}|\phi(w,A)|^3
\le \gamma_pB^3w^{p-3},
\]
and \(\gamma_p\) depends only on \(p\). Taking expectations and using the moments in \eqref{eqq4} and \eqref{eqq5}, we obtain
\[
\begin{aligned}
&\E[(w+\phi(w,A))^p-w^p]\\
&=pw^{p-1}\E[\phi(w,A)]\\
&\quad+\frac{p(p-1)}2w^{p-2}\E[\phi^2(w,A)]+O(w^{p-3})\\
&=p\left(K(y,x^\ast)+\frac{pV(y,x^\ast)}{2}\right)w^{p-2}
+O(w^{p-3}).
\end{aligned}
\]
This proves \eqref{eq:zeroH_reciprocal_power_drift}.
\end{proof}

By the bounds established in Appendix~\ref{app:strict_sign_invasion}, every fixed power of the stopped reciprocal odds is integrable at each finite time. Before exit, \(w_t>w_\varepsilon=1/z_\varepsilon\). Since \(w_\varepsilon\to\infty\) as \(\varepsilon\downarrow0\), the preceding estimates apply before exit for every sufficiently small \(\varepsilon\). Independence of \(A(t)\) from \(\mathcal F_t\) then gives the corresponding conditional estimates at \(w=w_t\).

\subsection{Proof of Result~\ref{thm:zeroH_positiveK}(i) (Attraction when \(K(y,x^\ast)>0\))}

Fix
\[
\begin{aligned}
&0<\theta<\frac{2K(y,x^\ast)}{V(y,x^\ast)},\\
&\kappa:=\frac{\theta}{2}\left(K(y,x^\ast)-\frac{\theta V(y,x^\ast)}{2}\right)>0.
\end{aligned}
\]
By Lemma~\ref{lem:zeroH_reciprocal} with \(p=-\theta\), we have
\[
\E_{q_0}[z_{t+1}^{\theta}-z_t^{\theta}\mid\mathcal F_t]
=-2\kappa z_t^{\theta+2}+O(z_t^{\theta+3}).
\]
Since the remainder is uniform and of higher order, choosing \(\varepsilon\) sufficiently small ensures that its absolute value is at most \(\kappa z_t^{\theta+2}\) whenever \(z_t<z_\varepsilon\).
Consequently,
\begin{equation}\label{eq:zeroH_drift_z}
\E_{q_0}[z_{t+1}^{\theta}\mid\mathcal F_t]
\le z_t^{\theta}-\kappa z_t^{\theta+2}
\quad\text{on }\{t<\tau_\varepsilon\}.
\end{equation}

Fix \(0<q_0<\varepsilon\). Applying Lemma~\ref{lem:stopped_drift_estimates}\textup{(i)} to \eqref{eq:zeroH_drift_z} with \(c=\kappa\) and \(\gamma=2\), we obtain
\begin{equation}\label{eq:zeroH_exit_bound}
\mathbb P_{q_0}(\tau_\varepsilon<\infty)
\le\left(\frac{z_0}{z_\varepsilon}\right)^\theta
\end{equation}
and \(q_t\to0\) a.s. on \(\{\tau_\varepsilon=\infty\}\). These conclusions hold for every sufficiently small \(\varepsilon\). By Remark~\ref{remark_appendixA_1}, resident fixation is locally asymptotically stable in probability against \(y\).

\subsection{Proof of Result~\ref{thm:zeroH_escape}\textup{(i)} (Escape when \(K(y,x^\ast)\le0\) and \(V(y,x^\ast)>0\))}
The proof uses two lemmas. The first shows that, on nonexit, the reciprocal odds \(w_t\) converge to a finite limit, so their increments tend to zero. The second shows that, on the same event, the absolute increments exceed a fixed positive constant infinitely often. Combining these conclusions rules out a positive probability of nonexit.

Following Lamperti's recurrence argument \cite{Lamperti1960}, define
\[
f(w):=\log\log(e+w),\qquad w\ge0.
\]
The function \(f\) is nonnegative, strictly increasing, and unbounded. This choice yields a negative drift for sufficiently large reciprocal odds even in the critical case \(K(y,x^\ast)=0\).

\begin{lemma}\label{lem:zeroH_convergence_on_nonexit}
Suppose that \(H(y,x^\ast)=0\), \(K(y,x^\ast)\le0\), and \(V(y,x^\ast)>0\). For every sufficiently small \(\varepsilon>0\) and every \(0<q_0<\varepsilon\), the process \((f(w_{t\wedge\tau_\varepsilon}))_{t\ge0}\) is a nonnegative supermartingale. On \(\{\tau_\varepsilon=\infty\}\), the reciprocal odds \(w_t\) converge to a finite limit a.s., in particular, \(w_{t+1}-w_t\to0\) a.s. on this event.
\end{lemma}

\begin{proof}
Taylor's theorem gives
\[
\begin{aligned}
&f(w+\phi(w,A))-f(w)\\
&=f'(w)\phi(w,A)+\frac12f''(w)\phi(w,A)^2+\frac16f'''(\zeta)\phi(w,A)^3,
\end{aligned}
\]
where \(\zeta\) lies between \(w\) and \(w+\phi(w,A)\). Differentiating \(f\), we obtain
\[
\begin{aligned}
f'(w)&=\frac{1}{(e+w)\log(e+w)},\\
f''(w)&=-\frac{\log(e+w)+1}{(e+w)^2[\log(e+w)]^2}.
\end{aligned}
\]
Moreover,
\[
\sup_{|u-w|\le B}|f'''(u)|
=O\!\left(\frac{1}{w^3\log w}\right)
\quad\text{as }w\to\infty.
\]
Since \(|\phi(w,A)|\le B\), the Taylor remainder is \(O(1/(w^3\log w))\), uniformly over environmental realizations.

Taking expectations and using \eqref{eq:zeroH_increment_moments}, we obtain
\begin{equation}\label{eq:zeroH_loglog_drift}
\begin{aligned}
&\E[f(w+\phi(w,A))-f(w)]\\
&=f'(w)\E[\phi(w,A)]+\frac12f''(w)\E[\phi(w,A)^2]
  +O\!\left(\frac{1}{w^3\log w}\right)\\
&=\frac{K(y,x^\ast)+V(y,x^\ast)/2}{w^2\log w}
  -\frac{V(y,x^\ast)(\log w+1)}{2w^2(\log w)^2}\\
 &\qquad \qquad \qquad \qquad 
  +O\!\left(\frac{1}{w^3\log w}\right)\\
&=\frac{K(y,x^\ast)}{w^2\log w}
  -\frac{V(y,x^\ast)}{2w^2(\log w)^2}
  +O\!\left(\frac{1}{w^3\log w}\right).
\end{aligned}
\end{equation}
In the second equality, replacing \(e+w\) by \(w\) contributes only to the stated remainder.

At \(K(y,x^\ast)=0\), the negative leading term dominates the remainder because \((\log w)/w\to0\). Since \(V(y,x^\ast)>0\), the absolute value of the remainder is at most \(V(y,x^\ast)/[4w^2(\log w)^2]\) for all sufficiently large \(w\). When \(K(y,x^\ast)<0\), the additional term \(K(y,x^\ast)/(w^2\log w)\) is nonpositive. Thus, in both cases, there exists \(W>e\) such that
\[
\E[f(w+\phi(w,A))-f(w)]
\le-\frac{V(y,x^\ast)}{4w^2(\log w)^2}<0,
\]
for $w\ge W$.

Choose \(\varepsilon>0\) sufficiently small that \(w_\varepsilon\ge W\), and fix \(0<q_0<\varepsilon\). Since \(w_t\) is \(\mathcal F_t\)-measurable and \(A(t)\) is independent of \(\mathcal F_t\), conditioning fixes \(w_t\) and averages over the current environment. On \(\{t<\tau_\varepsilon\}\), we have \(w_t>w_\varepsilon\ge W\), so
\[
\E_{q_0}[f(w_{t+1})-f(w_t)\mid\mathcal F_t]
\le-\frac{V(y,x^\ast)}{4w_t^2(\log w_t)^2}<0.
\]
The stopped process is constant after exit. Therefore, \((f(w_{t\wedge\tau_\varepsilon}))_{t\ge0}\) is a nonnegative supermartingale. It is integrable at each finite time because the bounded increments give
\[
w_{t\wedge\tau_\varepsilon}\le w_0+Bt.
\]
By the supermartingale convergence theorem, there exists a finite random variable \(L\ge0\) such that
\[
f(w_{t\wedge\tau_\varepsilon})\longrightarrow L
\quad\text{a.s.}
\]
On \(\{\tau_\varepsilon=\infty\}\), continuity of \(f^{-1}\) then gives
\[
w_t\longrightarrow f^{-1}(L)=\exp(\exp L)-e<\infty
\quad\text{a.s.}
\]
Consequently, \(w_{t+1}-w_t\to0\) a.s. on \(\{\tau_\varepsilon=\infty\}\).
\end{proof}

\begin{lemma}\label{lem:zeroH_persistent_increments}
Suppose that \(V(y,x^\ast)>0\). For every \(\varepsilon\in(0,1)\), there exists \(\eta_\varepsilon>0\) such that, for every \(0<q_0<\varepsilon\), the inequality \(\lvert w_{t+1}-w_t\rvert>\eta_\varepsilon\) holds infinitely often a.s. on \(\{\tau_\varepsilon=\infty\}\).
\end{lemma}
\begin{proof}
Before exit, \(w_t>w_\varepsilon\) and \(\alpha(t)\le M/m\). Hence
\[
|w_{t+1}-w_t|
=\frac{w_t|\beta(t)-\alpha(t)|}{w_t+\alpha(t)}
\ge c_\varepsilon|\beta(t)-\alpha(t)|,
\]
where $c_\varepsilon:=\dfrac{w_\varepsilon}{w_\varepsilon+M/m}>0$.
Since \(V(y,x^\ast)=\E[(\alpha-\beta)^2]>0\), there exists \(\delta>0\) such that
\[
\mathbb P(|\beta(0)-\alpha(0)|>\delta)>0.
\]
The events \(\{|\beta(t)-\alpha(t)|>\delta\}\), \(t\ge0\), are independent and have the same positive probability. By the second Borel--Cantelli lemma, they occur infinitely often a.s. On \(\{\tau_\varepsilon=\infty\}\), the increment bound holds at every generation. Taking \(\eta_\varepsilon=c_\varepsilon\delta\) therefore gives the claim.
\end{proof}

\begin{proof}[\textbf{Proof of Result~\ref{thm:zeroH_escape}\textup{(i)}}]
Fix \(\varepsilon>0\) sufficiently small for Lemma~\ref{lem:zeroH_convergence_on_nonexit} to apply, and let \(0<q_0<\varepsilon\). Suppose, for a contradiction, that \(\mathbb P_{q_0}(\tau_\varepsilon=\infty)>0\). Almost surely on this event, Lemma~\ref{lem:zeroH_convergence_on_nonexit} gives \(w_{t+1}-w_t\to0\), whereas Lemma~\ref{lem:zeroH_persistent_increments} gives \(\lvert w_{t+1}-w_t\rvert>\eta_\varepsilon\) infinitely often for some \(\eta_\varepsilon>0\). These conclusions are incompatible. Hence
\[
\mathbb P_{q_0}(\tau_\varepsilon<\infty)=1.
\]
This holds for every sufficiently small \(\varepsilon\) and every \(0<q_0<\varepsilon\). Thus resident fixation is not locally stable in probability against \(y\).
\end{proof}
\section{Proof of Result~\ref{thm:zeroH_positiveK}\textup{(ii)}}
\label{app:zeroH_rates}

Throughout this appendix, Assumptions \textup{($\mathcal A1$)}--\textup{($\mathcal A3$)} hold. Let \(x^\ast\) be an SNE and \(y\) a distinct tied mutant satisfying \(H(y,x^\ast)=0\) and \(K(y,x^\ast)>0\). 

We first establish moment bounds for the odds before exit and the stopped reciprocal odds. Using maximal inequalities and the Borel--Cantelli lemma, we then obtain upper and lower bounds on the growth of \(w_t\) on the event of nonexit. These bounds yield \(\log w_t/\log t\to1/2\). Finally, the identity \(q_t=1/(1+w_t)\) gives the logarithmic extinction exponent \(-1/2\).

\begin{lemma}
\label{lem:zeroH_moment_estimates}
Fix \(0<\theta<\frac{2K(y,x^\ast)}{V(y,x^\ast)}\), and set
\[
\begin{aligned}
\kappa&:=\frac{\theta}{2}\left(K(y,x^\ast)-\frac{\theta V(y,x^\ast)}{2}\right)>0,\\
\sigma&:=3\bigl(K(y,x^\ast)+V(y,x^\ast)\bigr)>0.
\end{aligned}
\]
For every sufficiently small \(\varepsilon\in(0,1)\) and every \(0<q_0<\varepsilon\), the process \((z_{t\wedge\tau_\varepsilon}^{\theta})_{t\ge0}\) is a nonnegative supermartingale, and \((w_{t\wedge\tau_\varepsilon}^{2})_{t\ge0}\) is a nonnegative submartingale. Moreover, for every integer \(t\ge0\),
\begin{subequations}\label{eq:zeroH_moment_estimates}
\begin{align}
\E_{q_0}\!\left[w_{t\wedge\tau_\varepsilon}^{2}\right]&\le w_0^2+\sigma t,\label{eq:zeroH_moment_estimates_2}\\
\E_{q_0}\!\left[z_t^\theta\mathbf1_{\{t<\tau_\varepsilon\}}\right]
&\le\left(z_0^{-2}+\frac{2\kappa}{\theta}t\right)^{-\theta/2}.\label{eq:zeroH_moment_estimates_1}
\end{align}
\end{subequations}
\end{lemma}

\begin{proof}
Inequality~\eqref{eq:zeroH_drift_z} together with \eqref{eq:stopped_increment_identity}, shows that \((z_{t\wedge\tau_\varepsilon}^{\theta})_{t\ge0}\) is a nonnegative supermartingale for sufficiently small \(\varepsilon\). In addition, applying Lemma~\ref{lem:zeroH_reciprocal} with \(p=2\) gives
\[
\E_{q_0}[w_{t+1}^2-w_t^2\mid\mathcal F_t]=2(K(y,x^\ast)+V(y,x^\ast))+O(z_t).
\]
Since \(K(y,x^\ast)+V(y,x^\ast)>0\), we can choose \(\varepsilon\) sufficiently small that the remainder has absolute value at most \(K(y,x^\ast)+V(y,x^\ast)\) on \(\{t<\tau_\varepsilon\}\). It follows that
\begin{equation}\label{eq:zeroH_drift_w2}
0\le
\E_{q_0}[w_{t+1}^2-w_t^2\mid\mathcal F_t]
\le \sigma 
\quad\text{on }\{t<\tau_\varepsilon\}.
\end{equation}
We also require \(\kappa z_\varepsilon^2<1\). The fitness and increment bounds ensure that all stopped random variables used below are integrable at every finite time.

By the identity \eqref{eq:stopped_increment_identity}, we have
\[
0\le
\E_{q_0}\!\left[
w_{(t+1)\wedge\tau_\varepsilon}^{2}
-w_{t\wedge\tau_\varepsilon}^{2}
\,\middle|\,\mathcal F_t
\right]
\le \sigma \mathbf1_{\{t<\tau_\varepsilon\}}.
\]
The lower bound shows that \((w_{t\wedge\tau_\varepsilon}^{2})_{t\ge0}\) is a submartingale. Taking expectations in the upper bound and summing from \(s=0\) to \(s=t-1\), we obtain
\[
\E_{q_0}[w_{t\wedge\tau_\varepsilon}^{2}]
\le w_0^2+\sigma \sum_{s=0}^{t-1}\mathbb P_{q_0}(s<\tau_\varepsilon)
\le w_0^2+\sigma t.
\]
This proves \eqref{eq:zeroH_moment_estimates_2}.

To prove \eqref{eq:zeroH_moment_estimates_1}, define
\[
u_t:=\E_{q_0}\!\left[z_t^\theta\mathbf1_{\{t<\tau_\varepsilon\}}\right].
\]
The indicator sets each path's contribution to zero after exit; this differs from stopping the odds at their exit value. The inclusion \(\{t+1<\tau_\varepsilon\}\subseteq\{t<\tau_\varepsilon\}\), the \(\mathcal F_t\)-measurability of \(\{t<\tau_\varepsilon\}\), and \eqref{eq:zeroH_drift_z} give
\begin{equation}
\begin{aligned}
u_{t+1}
&\le \E_{q_0}\!\left[z_{t+1}^{\theta}\mathbf1_{\{t<\tau_\varepsilon\}}\right]\\
&=\E_{q_0}\!\left[
\mathbf1_{\{t<\tau_\varepsilon\}}
\E_{q_0}[z_{t+1}^{\theta}\mid\mathcal F_t]
\right]\\
&\le u_t-\kappa
\E_{q_0}\!\left[z_t^{\theta+2}\mathbf1_{\{t<\tau_\varepsilon\}}\right].
\end{aligned}
\end{equation}
Since \(v\mapsto v^{1+2/\theta}\) is convex on \([0,\infty)\), Jensen's inequality gives
\[
\begin{aligned}
\E_{q_0}\!\left[z_t^{\theta+2}\mathbf1_{\{t<\tau_\varepsilon\}}\right]
&=\E_{q_0}\!\left[
\left(z_t^\theta\mathbf1_{\{t<\tau_\varepsilon\}}\right)^{1+2/\theta}
\right]\ge u_t^{1+2/\theta}.
\end{aligned}
\]
Consequently,
\begin{equation}\label{eq:zeroH_moment_recursion}
u_{t+1}\le u_t\bigl(1-\kappa u_t^{2/\theta}\bigr).
\end{equation}

Thus \((u_t)_{t\ge0}\) is nonnegative and nonincreasing. If \(u_t=0\) for some \(t\), all subsequent terms are zero and satisfy \eqref{eq:zeroH_moment_estimates_1}. Now suppose that \(u_{t+1}>0\). Then \(u_t>0\), and, since \(z_t<z_\varepsilon\) on \(\{t<\tau_\varepsilon\}\),
\[
u_t\le z_\varepsilon^\theta\mathbb P_{q_0}(t<\tau_\varepsilon)
\le z_\varepsilon^\theta.
\]
Therefore,
\[
0\le\kappa u_t^{2/\theta}
\le\kappa z_\varepsilon^2<1.
\]
Applying the inequality
\[
(1-v)^{-2/\theta}\ge1+\frac{2v}{\theta},
\quad 0\le v<1,
\]
to \eqref{eq:zeroH_moment_recursion}, we obtain
\[
\begin{aligned}
u_{t+1}^{-2/\theta}
&\ge u_t^{-2/\theta}
\bigl(1-\kappa u_t^{2/\theta}\bigr)^{-2/\theta}\\
&\ge u_t^{-2/\theta}+\frac{2\kappa}{\theta}.
\end{aligned}
\]
If \(u_t>0\), all preceding terms are positive. Iterating this inequality and using \(u_0=z_0^\theta\), we obtain
\[
u_t^{-2/\theta}
\ge z_0^{-2}+\frac{2\kappa}{\theta}t.
\]
Taking the power \(-\theta/2\) on both sides proves \eqref{eq:zeroH_moment_estimates_1}. As noted above, the same bound holds when \(u_t=0\).
\end{proof}

\begin{lemma}
\label{lem:zeroH_log_rate}
Under the conditions of Lemma~\ref{lem:zeroH_moment_estimates}, we obtain
\[
\lim_{t\to\infty}\frac{\log w_t}{\log t}
=\frac12
\quad\text{a.s. on }\{\tau_\varepsilon=\infty\}.
\]
\end{lemma}

\begin{proof}
Fix \(\theta\) and \(\varepsilon\) as in Lemma~\ref{lem:zeroH_moment_estimates}. Keep this choice of \(\varepsilon\) fixed as \(\delta\) varies. Let \(0<\delta<1/2\), and define \(t_j:=2^j\) for \(j\ge0\). We first prove an upper bound and then a lower bound for \(w_t\).

\textbf{\emph{Upper bound.}} Since \((w_{t\wedge\tau_\varepsilon}^{2})_{t\ge0}\) is a nonnegative submartingale, Doob's maximal inequality and \eqref{eq:zeroH_moment_estimates_2} give
\[
\begin{aligned}
\mathbb P_{q_0}\!\left(
\max_{0\le s\le t_{j+1}}w_{s\wedge\tau_\varepsilon}^{2}
\ge t_j^{1+2\delta}
\right)&\le
\frac{\E_{q_0}[w_{t_{j+1}\wedge\tau_\varepsilon}^{2}]}
{t_j^{1+2\delta}}\\
&\le
w_0^2\,2^{-j(1+2\delta)}
+2\sigma \,2^{-2j\delta}.
\end{aligned}
\]
The sum of the right-hand side over \(j\ge0\) is finite. By the first Borel--Cantelli lemma, there is a finite random integer \(J\), a.s., such that
\[
\max_{0\le s\le t_{j+1}}w_{s\wedge\tau_\varepsilon}^{2}
<t_j^{1+2\delta}
\quad\text{for every }j\ge J.
\]
On \(\{\tau_\varepsilon=\infty\}\), we have \(w_{s\wedge\tau_\varepsilon}=w_s\) for every \(s\ge0\). For every sufficiently large integer \(t\), choose \(j\ge J\) such that \(t_j\le t<t_{j+1}\). Then
\[
w_t<t_j^{1/2+\delta}\le t^{1/2+\delta}.
\]
Taking logarithms, dividing by \(\log t\), and letting \(t\to\infty\), we obtain
\begin{equation}\label{eq:zeroH_log_rate_upper}
\limsup_{t\to\infty}\frac{\log w_t}{\log t}
\le\frac12+\delta
\quad\text{a.s. on }\{\tau_\varepsilon=\infty\}.
\end{equation}

\textbf{\emph{Lower bound.}} For each fixed \(j\), define
\[
X_s^{(j)}
:=\mathbf1_{\{t_j<\tau_\varepsilon\}}
z_{s\wedge\tau_\varepsilon}^{\theta},
\quad s\ge t_j.
\]
The event \(\{t_j<\tau_\varepsilon\}\) is \(\mathcal F_{t_j}\)-measurable and hence \(\mathcal F_s\)-measurable for every \(s\ge t_j\). Multiplying the conditional supermartingale inequality for \(z_{s\wedge\tau_\varepsilon}^{\theta}\) by its indicator shows that \((X_s^{(j)})_{s\ge t_j}\) is a nonnegative supermartingale. Moreover, \eqref{eq:zeroH_moment_estimates_1} gives
\[
\E_{q_0}[X_{t_j}^{(j)}]
=\E_{q_0}\!\left[z_{t_j}^{\theta}\mathbf1_{\{t_j<\tau_\varepsilon\}}\right]
\le C_\theta t_j^{-\theta/2},
\]
where \(C_\theta:=\left(\theta/(2\kappa)\right)^{\theta/2}\). For every integer \(N\ge t_j\), the maximal inequality for nonnegative supermartingales yields
\[
\begin{aligned}
&\mathbb P_{q_0}\!\left(
t_j<\tau_\varepsilon,\;
\max_{t_j\le s\le N}z_{s\wedge\tau_\varepsilon}
>t_j^{-1/2+\delta}
\right)\\
&=
\mathbb P_{q_0}\!\left(
\max_{t_j\le s\le N}X_s^{(j)}
>t_j^{-\theta(1/2-\delta)}
\right)\\
&\quad\le
t_j^{\theta(1/2-\delta)}\E_{q_0}[X_{t_j}^{(j)}]
\le C_\theta t_j^{-\theta\delta}.
\end{aligned}
\]
The equality follows because \(\theta>0\) and \(X_s^{(j)}=z_{s\wedge\tau_\varepsilon}^{\theta}\) on \(\{t_j<\tau_\varepsilon\}\), while \(X_s^{(j)}=0\) otherwise.

As \(N\to\infty\), the events on the left increase to the event that the supremum over all \(s\ge t_j\) exceeds the same threshold. The monotone convergence theorem, applied to their indicators, therefore gives
\[
\mathbb P_{q_0}\!\left(
t_j<\tau_\varepsilon,\;
\sup_{s\ge t_j}z_{s\wedge\tau_\varepsilon}
>t_j^{-1/2+\delta}
\right)
\le C_\theta\,2^{-j\theta\delta}.
\]
Since the right-hand side is summable in \(j\), the first Borel--Cantelli lemma shows that, a.s. on \(\{\tau_\varepsilon=\infty\}\),
\[
z_s\le t_j^{-1/2+\delta}
\quad\text{for every }s\ge t_j
\]
for all sufficiently large \(j\). If \(t_j\le t<t_{j+1}=2t_j\), we therefore have
\[
w_t=\frac1{z_t}
\ge t_j^{1/2-\delta}
\ge(t/2)^{1/2-\delta}.
\]
The last inequality follows from \(t_j>t/2\) and \(1/2-\delta>0\). Taking logarithms, dividing by \(\log t\), and letting \(t\to\infty\), we obtain
\begin{equation}\label{eq:zeroH_log_rate_lower}
\liminf_{t\to\infty}\frac{\log w_t}{\log t}
\ge\frac12-\delta
\quad\text{a.s. on }\{\tau_\varepsilon=\infty\}.
\end{equation}

\textbf{\emph{Conclusion}.} Combining \eqref{eq:zeroH_log_rate_upper} and \eqref{eq:zeroH_log_rate_lower} gives
\[
\frac12-\delta
\le\liminf_{t\to\infty}\frac{\log w_t}{\log t}
\le\limsup_{t\to\infty}\frac{\log w_t}{\log t}
\le\frac12+\delta
\]
a.s. on \(\{\tau_\varepsilon=\infty\}\). These inequalities hold for every \(0<\delta<1/2\), so that 
$$\lim_{t\to\infty}\frac{\log w_t}{\log t}=\frac{1}{2}.$$
\end{proof}

\begin{proof}[\textbf{Proof of Result~\ref{thm:zeroH_positiveK}\textup{(ii)}}]
By Lemma~\ref{lem:zeroH_log_rate},
\[
\frac{\log w_t}{\log t}\longrightarrow\frac12
\quad\text{a.s. on }\{\tau_\varepsilon=\infty\}.
\]
In particular, \(w_t\to\infty\) on this event, outside a set of probability zero. Since \(q_t=1/(1+w_t)\), we have
\[
\log q_t=-\log w_t-\log(1+w_t^{-1}).
\]
The second term tends to zero. Dividing by \(\log t\) and using the preceding limit, we obtain
\[
\lim_{t\to\infty}\frac{\log q_t}{\log t}
=-\frac12
\quad\text{a.s. on }\{\tau_\varepsilon=\infty\}.
\]
This proves Result~\ref{thm:zeroH_positiveK}\textup{(ii)}.
\end{proof}

\section{Proof of Result~\ref{thm:zeroH_escape}\textup{(ii)}}
\label{appendix_proof_thm:zeroH_mean_exit}

Throughout this appendix, Assumptions \textup{($\mathcal A1$)}--\textup{($\mathcal A3$)} hold. Let \(x^\ast\) be an SNE and \(y\) a fixed distinct tied mutant satisfying \(H(y,x^\ast)=0\), \(K(y,x^\ast)\le0\), and \(V(y,x^\ast)>0\). We prove Result~\ref{thm:zeroH_escape}\textup{(ii)} by considering the three cases \(K(y,x^\ast)<-V(y,x^\ast)\), \(-V(y,x^\ast)<K(y,x^\ast)\le0\), and \(K(y,x^\ast)=-V(y,x^\ast)\) separately.

Since \(\E_{q_0}[w_{t+1}^2-w_t^2\mid\mathcal F_t]=2w_t\nu_1(w_t)+\nu_2(w_t)\), the moment expansions \eqref{eq:zeroH_increment_moments} give
\[
\E_{q_0}\!\left[w_{t+1}^2-w_t^2\mid\mathcal F_t\right]
=2\bigl(K(y,x^\ast)+V(y,x^\ast)\bigr)
+O(w_t^{-1}).
\]
We use this expansion in all three cases. For sufficiently small \(\varepsilon\), exit occurs a.s. by Result~\ref{thm:zeroH_escape}\textup{(i)}. It remains to determine whether its expectation is finite.

\subsection{Case 1: \(K(y,x^\ast)<-V(y,x^\ast)\)}\label{subsec:mean_exit_finite}
Set \(c:=-\bigl(K(y,x^\ast)+V(y,x^\ast)\bigr)>0\). Choose \(\varepsilon\) sufficiently small that the remainder in the quadratic drift has absolute value at most \(c\) on \(\{t<\tau_\varepsilon\}\). Then
\[
\E_{q_0}\!\left[w_{t+1}^2-w_t^2\mid\mathcal F_t\right]\le-c
\quad\text{on }\{t<\tau_\varepsilon\}.
\]
Applying Lemma~\ref{lem:stopped_drift_estimates}\textup{(ii)} with \(X_t=w_t^2\), we obtain
\[
\E_{q_0}[\tau_\varepsilon]\le\frac{w_0^2}{c}<\infty.
\]

\subsection{Case 2: \(-V(y,x^\ast)<K(y,x^\ast)\le0\)}
\label{subsec:mean_exit_infinite}

Choose
\[
-\frac{2K(y,x^\ast)}{V(y,x^\ast)}<p<2.
\]
This choice gives \(p>0\) and \(K(y,x^\ast)+pV(y,x^\ast)/2>0\). By Lemma~\ref{lem:zeroH_reciprocal},
\[
\begin{aligned}
&\E_{q_0}\!\left[w_{t+1}^p-w_t^p\mid\mathcal F_t\right]\\
&=p\left(K(y,x^\ast)+\frac{pV(y,x^\ast)}{2}\right)w_t^{p-2}
+O(w_t^{p-3}).
\end{aligned}
\]
Since the leading coefficient is positive, the conditional drift is nonnegative on \(\{t<\tau_\varepsilon\}\) for sufficiently small \(\varepsilon\). Therefore, \((w_{t\wedge\tau_\varepsilon}^p)_{t\ge0}\) is a submartingale, and
\[
\E_{q_0}[w_{t\wedge\tau_\varepsilon}^p]\ge w_0^p
\quad\text{for every }t\ge0.
\]
The expansion of the quadratic drift also gives a constant \(C_0>0\) such that
\begin{equation}\label{eq:mean_exit_quadratic_upper_bound}
\E_{q_0}\!\left[w_{t+1}^2-w_t^2\mid\mathcal F_t\right]\le C_0
\quad\text{on }\{t<\tau_\varepsilon\}.
\end{equation}

Suppose, for a contradiction, that \(\E_{q_0}[\tau_\varepsilon]<\infty\). Applying \eqref{eq:stopped_increment_identity} to \eqref{eq:mean_exit_quadratic_upper_bound}, taking expectations, and summing from \(s=0\) to \(s=t-1\), we obtain
\begin{equation}\label{eq:zeroH_L2_bound}
\begin{aligned}
\E_{q_0}[w_{t\wedge\tau_\varepsilon}^2]
&\le w_0^2+C_0\E_{q_0}[t\wedge\tau_\varepsilon]\le w_0^2+C_0\E_{q_0}[\tau_\varepsilon].
\end{aligned}
\end{equation}
Since \(2/p>1\), this gives
\[
\sup_{t\ge0}
\E_{q_0}\!\left[
\left(w_{t\wedge\tau_\varepsilon}^p\right)^{2/p}
\right]<\infty.
\]
Hence the family \(\{w_{t\wedge\tau_\varepsilon}^p:t\ge0\}\) is uniformly integrable. Moreover, \(\E_{q_0}[\tau_\varepsilon]<\infty\) implies \(\tau_\varepsilon<\infty\) a.s., and therefore
\[
w_{t\wedge\tau_\varepsilon}^p
\longrightarrow w_{\tau_\varepsilon}^p
\quad\text{a.s.}
\]
The a.s. convergence and uniform integrability imply convergence in \(L^1\), so we may pass to the limit in the expectation inequality. Since \(w_{\tau_\varepsilon}\le w_\varepsilon<w_0\), we obtain
\[
w_0^p
\le\E_{q_0}[w_{\tau_\varepsilon}^p]
\le w_\varepsilon^p
<w_0^p.
\]
This contradiction proves that \(\E_{q_0}[\tau_\varepsilon]=\infty\).

\subsection{Case 3: \(K(y,x^\ast)=-V(y,x^\ast)\)}
\label{subsec:mean_exit_critical}
At the critical threshold, the leading term in the drift of \(w_t^2\) vanishes. We first show that subtracting a suitable linear term from \(w_t^2\) gives a submartingale up to exit. We then show that a finite mean exit time would allow us to pass to the limit in its expectation inequality, contradicting the fact that the reciprocal odds are smaller at exit than initially. Write \(\xi_1(w):=\E[\phi(w,A)]\).

\begin{lemma}\label{lem:critical_exit_submartingale}
There exists \(C>0\) such that, for every sufficiently small \(\varepsilon>0\) and every \(0<q_0<\varepsilon\), the following hold.
\begin{enumerate}
\item[(i)] Almost surely, \(w_{t\wedge\tau_\varepsilon}>C\) for all \(t\ge0\). Moreover, \(\xi_1(w_t)<0\) on \(\{t<\tau_\varepsilon\}\).
\item[(ii)] The function \(\varphi(w)=w^2-Cw\) is strictly increasing on \((C,\infty)\), and \((\varphi(w_{t\wedge\tau_\varepsilon}))_{t\ge0}\) is a nonnegative submartingale.
\end{enumerate}
\end{lemma}

\begin{proof}
By Lemma~\ref{lem:zeroH_reciprocal}, we have
\[
\begin{aligned}
&\xi_1(w)=-\frac{V(y,x^\ast)}{2w}+O(w^{-2}),\\
&\E\!\left[(w+\phi(w,A))^2-w^2\right]=O(w^{-1}).
\end{aligned}
\]
It follows that there are constants \(W,L>0\) such that, for every \(w\ge W\),
\[
\begin{aligned}
&\xi_1(w)\le-\frac{V(y,x^\ast)}{4w},\\
&\E\!\left[(w+\phi(w,A))^2-w^2\right]\ge-\frac{L}{w}.
\end{aligned}
\]
Choose \(C>4L/V(y,x^\ast)\), and define \(\varphi(w):=w^2-Cw\). For \(w\ge W\), we obtain
\[
\begin{aligned}
&\E\!\left[\varphi(w+\phi(w,A))-\varphi(w)\right]\\
&=\E\!\left[(w+\phi(w,A))^2-w^2\right]-C\xi_1(w)\\
&\ge\frac{CV(y,x^\ast)/4-L}{w}>0.
\end{aligned}
\]
Choose \(\varepsilon\) sufficiently small that \(w_\varepsilon\ge W\) and \(w_\varepsilon>C+B\). Before exit, \(w_t>w_\varepsilon\), so \(\xi_1(w_t)<0\). Since an increment can decrease \(w_t\) by at most \(B\), we also have
\[
w_{t\wedge\tau_\varepsilon}>w_\varepsilon-B>C
\quad\text{for every }t\ge0.
\]
The function \(\varphi\) is nonnegative and strictly increasing on this range. Together with its positive conditional drift before exit and integrability at each finite time, this shows that \((\varphi(w_{t\wedge\tau_\varepsilon}))_{t\ge0}\) is a nonnegative submartingale.
\end{proof}

\begin{lemma}\label{lem:critical_exit_domination}
Let \(\varepsilon\) and \(q_0\) be as in Lemma~\ref{lem:critical_exit_submartingale}. If \(\E_{q_0}[\tau_\varepsilon]<\infty\), then
\[
\E_{q_0}\!\left[\sup_{t\ge0}w_{t\wedge\tau_\varepsilon}^{\,2}\right]<\infty.
\]
\end{lemma}
\begin{proof}
Assume that \(\E_{q_0}[\tau_\varepsilon]<\infty\). Define
\[
M_t:=
\sum_{s=0}^{t-1}\mathbf1_{\{s<\tau_\varepsilon\}}
\left(w_{s+1}-w_s-\xi_1(w_s)\right),
\]
with \(M_0=0\), and write \(\Delta M_s:=M_{s+1}-M_s\) for the increment at time \(s\). Since \(w_s\) is \(\mathcal F_s\)-measurable and \(A(s)\) is independent of \(\mathcal F_s\), we have
\[
\begin{aligned}
\mathbb E_{q_0}\left[w_{s+1}-w_s\mid\mathcal F_s\right]
&=\mathbb E_{q_0}[\phi(w_s,A(s))\mid\mathcal F_s]\\
&=\left.\mathbb E[\phi(w,A(s))]\right|_{w=w_s}=\xi_1(w_s).
\end{aligned}
\]
It follows that
\[
\begin{aligned}
\mathbb E_{q_0}[\Delta M_s\mid\mathcal F_s]&=\mathbf1_{\{s<\tau_\varepsilon\}}
\Bigl(\mathbb E_{q_0}[w_{s+1}-w_s\mid\mathcal F_s]-\xi_1(w_s)\Bigr)\\
&=\mathbf1_{\{s<\tau_\varepsilon\}}\bigl(\xi_1(w_s)-\xi_1(w_s)\bigr)=0.
\end{aligned}
\]
Thus \((M_t)_{t\ge0}\) is a martingale. The bounds \(|\phi|\le B\) and \(|\xi_1|\le B\) imply that \(|\Delta M_s|\le 2B\). For \(0\le r<s\), the increments are orthogonal as
\[
\E_{q_0}[\Delta M_r\,\Delta M_s]
=\E_{q_0}\!\left[\Delta M_r\,\E_{q_0}[\Delta M_s\mid\mathcal F_s]\right]
=0.
\]
Expanding the square and using this orthogonality, we obtain
\[
\begin{aligned}
\E_{q_0}[M_t^2]&=\E_{q_0}\Big[\Big(\sum_{s=0}^{t-1}\Delta M_s\Big)^2\Big]\\
&=\sum_{s=0}^{t-1}\E_{q_0}[(\Delta M_s)^2]+2\sum_{0\le r<s<t}\E_{q_0}[\Delta M_r\,\Delta M_s]\\
&=\sum_{s=0}^{t-1}\E_{q_0}[(\Delta M_s)^2]\\
&=\sum_{s=0}^{t-1}\E_{q_0}\!\left[\mathbf1_{\{s<\tau_\varepsilon\}}\left(w_{s+1}-w_s-\xi_1(w_s)\right)^2\right]\\
&\le 4B^2\sum_{s=0}^{t-1}\mathbb P_{q_0}(s<\tau_\varepsilon)\\
&=4B^2\E_{q_0}[t\wedge\tau_\varepsilon]\le 4B^2\E_{q_0}[\tau_\varepsilon].
\end{aligned}
\]
Combining Doob's \(L^2\) maximal inequality with this bound gives, for every integer \(N\ge0\),
\[
\E_{q_0}\!\left[\max_{0\le t\le N}|M_t|^2\right]
\le4\E_{q_0}[M_N^2]
\le16B^2\E_{q_0}[\tau_\varepsilon].
\]
Letting \(N\to\infty\) and applying the monotone convergence theorem, we obtain
\begin{equation}\label{eq:critical_mean_exit_martingale_bound}
\E_{q_0}\!\left[\sup_{t\ge0}|M_t|^2\right]
\le16B^2\E_{q_0}[\tau_\varepsilon]<\infty.
\end{equation}

By the definition of \(M_t\), the stopped reciprocal odds satisfy
\[
w_{t\wedge\tau_\varepsilon}
=w_0+M_t+
\sum_{s=0}^{t-1}\mathbf1_{\{s<\tau_\varepsilon\}}\xi_1(w_s).
\]
The last sum is nonpositive by Lemma~\ref{lem:critical_exit_submartingale}. Thus, for every \(t\),
\[
0<w_{t\wedge\tau_\varepsilon}\le w_0+M_t\le w_0+\sup_{s\ge0}|M_s|.
\]
Squaring and taking the supremum over \(t\) gives
\begin{equation}\label{eqq1}
\sup_{t\ge0}w_{t\wedge\tau_\varepsilon}^{\,2}\le\left(w_0+\sup_{t\ge0}|M_t|\right)^2.
\end{equation}
Using \((a+b)^2\le2a^2+2b^2\) and the martingale bound \eqref{eq:critical_mean_exit_martingale_bound}, we obtain
\begin{equation}\label{eqq2}
\begin{aligned}
\mathbb E_{q_0}\!\left[\left(w_0+\sup_{s\ge0}|M_s|\right)^2\right]
&\le 2w_0^2+2\mathbb E_{q_0}\!\left[\sup_{s\ge0}|M_s|^2\right]\\
&\le 2w_0^2+32B^2\mathbb E_{q_0}[\tau_\varepsilon]<\infty.
\end{aligned}
\end{equation}
Combining Eqs. \eqref{eqq1} and \eqref{eqq2} completes the proof.
\end{proof}

\begin{proof}[\textbf{Proof of Result~\ref{thm:zeroH_escape}\textup{(ii)} at the critical threshold}]
Fix \(\varepsilon\) and \(q_0\) as in Lemma~\ref{lem:critical_exit_submartingale}, and suppose, for a contradiction, that \(\E_{q_0}[\tau_\varepsilon]<\infty\). Then \(\tau_\varepsilon<\infty\) a.s., so \(\varphi(w_{t\wedge\tau_\varepsilon})\to\varphi(w_{\tau_\varepsilon})\) a.s. For every \(t\ge0\), we have
\[
0\le\varphi(w_{t\wedge\tau_\varepsilon})=w_{t\wedge\tau_\varepsilon}^2-Cw_{t\wedge\tau_\varepsilon}
\le\sup_{s\ge0}w_{s\wedge\tau_\varepsilon}^{\,2}.
\]
The upper bound is integrable by Lemma~\ref{lem:critical_exit_domination}. The dominated convergence theorem therefore gives the equality in
\begin{equation}\label{eqqqqq}
\varphi(w_0)
\le\lim_{t\to\infty}\E_{q_0}[\varphi(w_{t\wedge\tau_\varepsilon})]
=\E_{q_0}[\varphi(w_{\tau_\varepsilon})].
\end{equation}
The inequality follows because \((\varphi(w_{t\wedge\tau_\varepsilon}))_{t\ge0}\) is a submartingale by Lemma~\ref{lem:critical_exit_submartingale}\textup{(ii)}. On the other hand, Lemma~\ref{lem:critical_exit_submartingale}\textup{(i)} gives \(w_{\tau_\varepsilon}>C\) a.s., and the definition of exit gives \(w_{\tau_\varepsilon}\le w_\varepsilon<w_0\). Since \(\varphi\) is strictly increasing on \((C,\infty)\) by Lemma~\ref{lem:critical_exit_submartingale}\textup{(ii)}, we obtain
\begin{equation}
\E_{q_0}[\varphi(w_{\tau_\varepsilon})]
\le\varphi(w_\varepsilon)
<\varphi(w_0),
\end{equation}
which contradicts \eqref{eqqqqq}. Hence \(\E_{q_0}[\tau_\varepsilon]=\infty\).
\end{proof}

\section{Directional criterion and nearby mutants (Result~\ref{prop:SES_directional_refinement})}\label{app:SES_consequences}

Fix a nonzero \(h\in\mathcal N(x^\ast)\). For every sufficiently small \(s>0\), the strategy \(y_s=x^\ast+sh\) belongs to the simplex. Since
\[
y_s^{\T}A(t)x^\ast
=(x^\ast)^{\T}A(t)x^\ast
+s h^{\T}A(t)x^\ast
=c(t)
\quad\text{a.s.},
\]
tie rigidity implies that \(y_s\) is a tied best reply whenever it is feasible.

Using the definitions of \(B_h\) and \(C_h\), the normalized fitnesses satisfy
\[
\begin{aligned}
\beta_s
&=\frac{(x^\ast)^{\T}A(t)y_s}{c(t)}
=1+sB_h,\\
\alpha_s
&=\frac{y_s^{\T}A(t)y_s}{c(t)}
=1+sB_h+s^2C_h.
\end{aligned}
\]
Assumption \textup{($\mathcal A2$)} ensures that \(B_h\) and \(C_h\) are bounded for this fixed direction, so all expectations below are finite. In particular,
\[
\begin{aligned}
H(y_s,x^\ast)&=\E\left[\beta_s-\alpha_s\right]=-s^2\E[C_h],\\
V(y_s,x^\ast)&=\E[(\alpha_s-\beta_s)^2]=s^4\E[C_h^2].
\end{aligned}
\]
Moreover,
\[
\frac12(\alpha_s^2-\beta_s^2)
=s^2C_h+s^3B_hC_h+\frac{s^4}{2}C_h^2.
\]
When \(\E[C_h]=0\), taking expectations gives
\[
K(y_s,x^\ast)
=s^3\E[B_hC_h]+\frac{s^4}{2}\E[C_h^2].
\]
This proves the identities in \eqref{eq:directional_HK_scaling}.

\textbf{\emph{Necessity.}} Suppose first that \(x^\ast\) is SES. If \(\E[C_h]>0\), then \(H(y_s,x^\ast)<0\) for every feasible \(s>0\), contradicting Result~\ref{cor:zeroH_SES}. Therefore, \(\E[C_h]\le0\).

If \(\E[C_h]<0\), the first alternative in \eqref{eq:SES_directional_refinement} holds. Suppose instead that \(\E[C_h]=0\), which implies $H(y_s,x^\ast)=0$. If \(\E[B_hC_h]<0\), then
\[
K(y_s,x^\ast)
=s^3\left(\E[B_hC_h]+\frac{s}{2}\E[C_h^2]\right)<0
\]
for all sufficiently small \(s>0\). This again contradicts Result~\ref{cor:zeroH_SES}, so \(\E[B_hC_h]\ge0\). We must also have \(\E[C_h^2]>0\). Otherwise \(C_h=0\) a.s., giving \(\alpha_s=\beta_s\) a.s. and exact neutrality against every feasible \(y_s\) (Result \ref{res:exact_neutrality}). Such a mutant cannot satisfy the attraction requirement in the definition of SES. This proves necessity.

\textbf{\emph{Sufficiency.}} Conversely, suppose that every nonzero \(h\in\mathcal N(x^\ast)\) satisfies \eqref{eq:SES_directional_refinement}. Take any distinct tied best reply \(y\), and set \(h=y-x^\ast\). Then \(h\in\mathcal N(x^\ast)\setminus\{0\}\), and \(y=x^\ast+h\) corresponds to \(s=1\). If \(\E[C_h]<0\), then \(H(y,x^\ast)>0\). Under the second alternative,
\[
\begin{aligned}
H(y,x^\ast)&=0,\\
K(y,x^\ast)&=\E[B_hC_h]+\frac12\E[C_h^2]>0.
\end{aligned}
\]
Thus every distinct tied best reply satisfies \eqref{eq:zeroH_exact_SES}, and Result~\ref{cor:zeroH_SES} implies that \(x^\ast\) is SES.

\section{An SES with an intermediate best-reply set}\label{ex:intermediate_best_replies}

We extend \eqref{eq:fully_weak_SES_matrix} to a resident supported on a face. Let \(S=\{1,\ldots,k\}\), where \(2\le k\le n\), and define
\[
x^\ast=\frac1k\1_S,
\quad
P_S=\operatorname{diag}(\1_S)-\frac1k\1_S\1_S^{\T}.
\]
Consider
\begin{equation}\label{eq:noise_induced_SES_matrix}
A(t)=(\1+\1_S)\1^{\T}+\xi_tP_S,
\end{equation}
where the independent variables \(\xi_t\) take the values \(\pm r\) with equal probabilities and \(0<r<1\). The entries lie in \([1,3]\). When \(k=n\), this is the construction in \eqref{eq:fully_weak_SES_matrix}.

Since \(P_Sx^\ast=0\), we have \(A(t)x^\ast=\1+\1_S\) and \((x^\ast)^{\T}A(t)x^\ast=2\). Therefore,
\[
Q(y,x^\ast)
=\log\!\left(1-\frac12\sum_{j\notin S}y_j\right).
\]
It follows that \(x^\ast\) is an SNE and \(\BR(x^\ast)=\Delta_S\), where \(\Delta_S\) is the face of strategies supported on \(S\). The best-reply set is intermediate when \(k<n\) and fully weak when \(k=n\).

For a distinct tied mutant \(y\in\Delta_S\), put \(d(y):=\|y-x^\ast\|^2\). The identities \((x^\ast)^{\T}P_S=0\) and \(y^{\T}P_Sy=d(y)\) give
\[
\beta(t)=1,
\quad
\alpha(t)=1+\frac{\xi_td(y)}2.
\]
These expressions yield the coefficients in \eqref{eq:noise_induced_SES_coefficients}. Every distinct tied mutant is therefore resisted. Mutants outside \(\Delta_S\) have negative invasion exponents, so \(x^\ast\) is SES.

To prove uniqueness when \(k=n\), take any \(x\ne x^\ast\) and write \(d_x:=\|x-x^\ast\|^2\). Then \((x^\ast)^{\T}A(t)x=2\) and \(x^{\T}A(t)x=2+\xi_td_x\), giving
\[
Q(x^\ast,x)
=-\frac12\log\!\left(1-\frac{r^2d_x^2}{4}\right)>0.
\]
Here \(0<d_x\le1-1/n\), so the logarithmic argument lies in \((0,1)\). Thus every other resident has a mutant with positive invasion exponent and cannot be an SNE.

We next prove convergence from every deterministic \(q_0\in(0,1)\) against a fixed distinct tied mutant. The exact recursion becomes
\[
z_{t+1}=z_t\left(1+\frac{\xi_td(y)}2q_t\right),
\]
from which we have $\E_{q_0}[z_{t+1}\mid\mathcal F_t]=z_t$. By the martingale convergence theorem, the nonnegative martingale \((z_t)_{t\ge0}\) converges a.s. to a finite random variable \(L\). On the event \(\{L>0\}\),
\[
|z_{t+1}-z_t|
=\frac{rd(y)}2z_tq_t
\longrightarrow
\frac{rd(y)L^2}{2(1+L)}>0.
\]
This contradicts convergence of \(z_t\), which requires its increments to tend to zero. Hence \(L=0\) a.s., and \(q_t\to0\) a.s. In particular, when \(k=n\), this conclusion holds against every distinct mutant.

\bibliographystyle{apsrev4-2}
\bibliography{bibliography}
\end{document}